\documentclass[aps,prc,preprint,showpacs,showkeys,superscriptaddress]{revtex4-1}
\usepackage{amsfonts}
\usepackage{amsmath}
\usepackage{amssymb}
\usepackage{graphicx}
\usepackage{xcolor}
\usepackage[version=3]{mhchem}
\usepackage{url}
\usepackage{natbib}
\usepackage[colorlinks,urlcolor=blue,citecolor=blue]{hyperref}

\begin{document}
\title{Review of Gamow-Teller and Fermi Transition Strength Functions}
\author{Muna Al-Harby}
\affiliation{Physics Department, College of Science, Qassim University, Qassim, Burydah 51411, KSA. }

\author{Bassam Shehadeh}
\affiliation{Physics Department, College of Science, Qassim University, Qassim, Burydah 51411, KSA. }


\begin{abstract}
We studied the temperature effect in isospin-singlet pairings in Gamow-Teller excitations.
We use theories of a hole-particle in the mean field shell model studied decay transition
using the one-particle-one-hole model for the $\beta$-decay of odd-even isotopes and the
two-particle-hole models for the $\beta$-decay of even-even and/or odd-odd isotopes.
Our reference isotopes for the one-particle-one-hole model are \ce{^{15}O}, \ce{^{15}N}, \ce{^{17}F}, and \ce{^{41}Sc},
whereas for the two-particle-hole model we use \ce{^{16}N} (for $\beta^-$-decay)
and \ce{^{56}Ni} and \ce{^{40}Sc} (for $\beta^+$/EC).

The calculations involve evaluating the matrix elements of Gamow -Teller and Fermi transitions,
then calculate the reduced transition probabilities of Gamow-Teller and Fermi, from which we evaluate
the half-lives and the strength function $ft$. The results are compared with the available experimental data.
For one-particle-one-hole model we found there is a deviation from experimental values which indicates that the
model is not valid for beta decay for the even-even nuclei in the ground state due to the residual nucleon-nucleon
interaction. As for a two-particle-hole model, we calculated the transition amplitude, from which we calculated
the strength of the transition $\log ft$ values. We found an excellent agreement between experimental and theoretical results.

By drawing the relationship between temperature versus $\log ft$ values, we found the general trend is that the strength function values slowly
decrease as temperatures increases. There are fluctuations $\log ft$ due to the strongly dependent of $\log ft$ on the shell configuration of the valence nucleons.

\end{abstract}

\volumeyear{year}
\volumenumber{number}
\issuenumber{number}
\eid{identifier}
\date[Date text]{date}
\received[Received text]{date}
\revised[Revised text]{date}
\accepted[Accepted text]{date}
\published[Published text]{date}
\maketitle

\section{Introduction}
Nuclear $\beta$-decay plays an important role in nuclear physics in particular and in various branches of science in general,  such as astrophysics and particle physics. The investigation of $beta$-decay provides valuable insights into how effective nuclear interactions depend on the spin and isospin, as well as on nuclear properties such as masses, shapes, and level densities \cite{RevModPhys.75.1021,PhysRevLett.101.142504}. In astrophysics, $\beta$-decay is responsible for the formation of neutron stars, the factories of heavy elements in our universe \cite{Qian_2008} by setting the time scale of the rapid neutron-capture via the half-lives of $\beta$-decay. In particle physics, $\beta$-decay offers the first experimental evidence of parity violation \cite{WuPhysRev57}, and is utilized to verify the unitarity of the Cabibbo–Kobayashi–Maskawa (CKM) matrix \cite{Towner_2010}.

Beta decay results from the presence of a weak force, which undergoes
a relatively slow decay time. Nucleons are formed from up and down quarks, and the quark has a weak strength which contributes to changing the flavor of lepton by producing the $W$ boson, which produces an electron/antineutrino or positron/neutrino pair. The most obvious example is the decay of a neutron, which consists of an up quark and two down quarks, to produce a proton that consists of two up quarks and a down quark.

The process of measuring the beta decay strength functions, and hence the decay half-lives, in an accurate way, is a required and necessary method and can be applied Beyond the Standard Model (BSM) as a way
to search for new fundamental physics can be discovered through beta
decay in atomic nuclei. The problem of understanding the 'background' of the Standard Model and its related perception of the low-energy quantum chromodynamic effects that appear in the form of a nuclear structure is one of the outstanding difficulties hampering the discovery of
new physics.

The difficulty of treating experimentally-favored nuclei theoretically in
a framework that allows for measured uncertainty makes the issue worse.
Nevertheless, advancements have been achieved over the past several decades that allow for the systematic construction of the internucleon interaction within a strong field theory framework. The medium-mass
nuclei, frequently relevant for BSM searches, can now be treated ab initio, thanks to developments in the many-body theory and computer power \cite{PhysRevC.99.064307}. Of course, there is still much to be learned about effective-field theory and how approximation strategies used in ab initio computations affect the relevant observables.

With the achieved advances in the measurement of nuclear $\beta$-decay half-lives due to the development of radioactive ion-beam facilities, a complete listing of the experimental data is available nowadays \cite{Gove1971}. It is imperative to test the fundamental theoretical model and their abilities to reproduce the experimental values to determine the weak and/or strength points in these models.

This review study presents the latest conventions in the theory of $\beta$-decay theory and procedures to evaluate the transition matrices due to the $\beta$-decay \cite{Suhonen2017}. Then focuses on $\beta$-decay transition in one-particle-hole and two-particle-hole nuclei schemes. The one-particle-hole theory is utilized to predict the half-lives of odd-even nuclei: \ce{^{15}O}, \ce{^{17}F}, \ce{^{39}Ca}, and \ce{^{41}Sc}. whereas, the two-particle-hole theory is utilized to calculate the strength functions of the EC/$\beta^+$-decay of \ce{^{56}Ni} as an even-even nucleus, the $\beta$-decay of \ce{^{16}N} as an odd-odd nucleus, and finally the EC/$\beta^+$-decay of \ce{^{40}Sc} as an odd-odd nucleus. The reason for choosing these isotopes is to study the effect of residual NN force on the valence nucleons. It is expected that such a force plays a greater role for an even number of nucleons than an odd number \cite{Shirokov2005} which suppresses the single particle transitions. 
The temperature effect on the strength functions is presented. Finally, the study presents conclusions and suggestions for further studies.
\section{Theoretical Background}
\subsection{Theory of Nuclear $\beta$-decay}

\begin{figure}[ptbh]
\begin{center}
\includegraphics[scale=0.4]{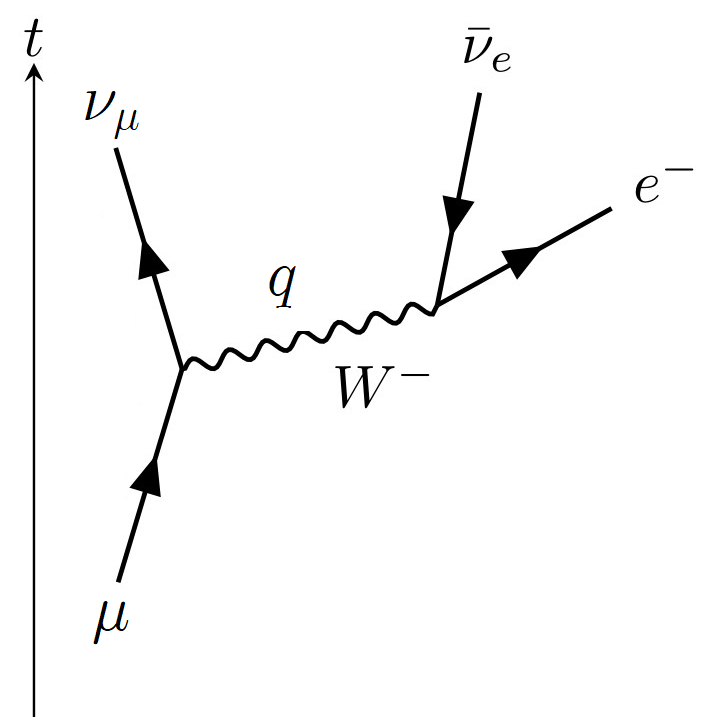}
\end{center}
\par\vspace{-0.7cm}
\caption{Feynman diagram depicts the
weak muon decay. The $W$-boson propagator carries momentum $q$, where $%
q^{2}\ll (m_{W}c^{2})^{2}$, for precise measurements of the weak coupling $%
g_{W}$.}
\label{fig_muon_decay}
\end{figure}

\begin{figure}[ptbh]
\begin{center}
\includegraphics[scale=0.4]{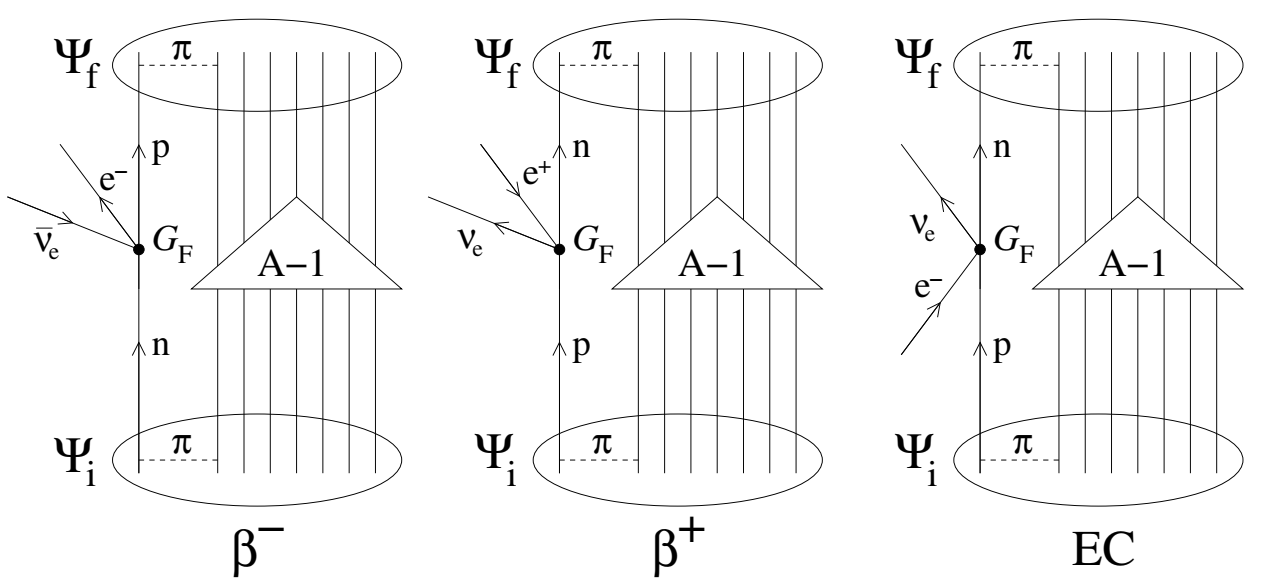}
\end{center}
\par\vspace{-0.7cm}
\caption{Nuclear $\beta^-$, $\beta^+$, and EC
decay in the impulse approximation. In this picture only one nucleon
contributes in the $\beta$ decay process whereas the
remaining $A-1$ nucleons are spectators. The initial and final states
$\Psi_i$ and $\Psi_f$ are the intial and final nuclear states of a strongly
interacting $A$-body wave function. At the weak-interaction vertices
the antilepton lines are drawn as going backwards in time. The strength
of the pointlike effective weak interaction vertex is given by the
Fermi constant $G_F$.}
\label{fig_NuclearBeta}
\end{figure}

In the nuclear scale the $\beta ^{-}$-decay is written as
\begin{equation}
_{_{Z}}^{^{A}}\text{X}_{_{N}}\rightarrow _{_{Z+1}}^{^{A}}\text{Y}%
_{_{N-1}}+e^{-}+\bar{\nu}_{e}.  \label{eq_nbetaminus}
\end{equation}%
The nuclear $\beta ^{+}$-decay is%
\begin{equation}
_{_{Z}}^{^{A}}\text{X}_{_{N}}\rightarrow _{_{Z-1}}^{A}\text{Y}%
_{_{N+1}}+e^{+}+\nu _{e}.  \label{eq_nbetaplus}
\end{equation}%
Finally nuclear EC reads%
\begin{equation}
_{_{Z}}^{^{A}}\text{X}_{_{N}}+e^{-}\rightarrow _{_{Z-1}}^{^{A}}\text{Y}%
_{_{N+1}}+\bar{\nu}_{e}.  \label{eq_nEC}
\end{equation}
In the three processes, shown in fig.(\ref{fig_NuclearBeta}), the parent
nucleus $^{A}$X and the daughter nucleus $^{A}$Y are isobars, i.e. both have
the same mass number $A$. This process has a coupling constant $G_{F}$ which
is not fundamental. It involves two fundamental vertices of weak coupling $%
g_{W}$. Strength of weak interaction is measured in muon decay, shown in
fig.(\ref{fig_muon_decay}), where $q^{2}<m_{\mu }c^{2}=106\,\mathrm{MeV}$.
Thus the $W$-boson propagator in natural unit can be written 
\begin{equation*}
\frac{-i\left( g_{\mu \nu }-q_{\mu }q_{\nu }/m_{W}^{2}\right) }{%
q^{2}-m_{W}^{2}}\approx \frac{ig_{\mu \nu }}{m_{W}^{2}}.
\end{equation*}%
In muon decay this becomes $g_{W}^{2}/m_{W}^{2}$. Hence, the weak coupling $%
g_{W}$ can be related to $G_{F}$ using \cite{halzen84} 
\begin{equation}
\frac{G_{F}}{\sqrt{2}}=\frac{g_{W}^{2}}{8(m_{w}c^{2})^{2}}.  \label{eq64}
\end{equation}%
This is valid for large mass of $W$-boson and small energy $q^{2}$ of $\beta 
$-decay, i.e. $q^{2}\ll (m_{W}c^{2})^{2}$. In case of $q^{2}\geq
(m_{W}c^{2})^{2}$ the weak interaction is more probable than electromagnetic
force. In another word, weak interaction is only weak because of the large $W
$-boson mass ($m_{W}=80.403\pm 0.029$ GeV/$c^{2}$). For muon decay $%
G_{F}=1.16639(1)\times 10^{-5}$ GeV$^{-2}$ \cite{halzen84}.

\subsection{Allowed Beta Decay}

We consider leptons final state emitted in $s$-state $(l=0)$ relative to the
nucleus (isotropic emission). Similarly, in allowed EC the initial electron
is from an $s$-shell, and the final neutrino is in an $s$-state relative to
the nucleus. Other $\beta $-decay process involves higher values of leptons
orbital angular moment ($p$-state, $d$-state etc.) are traditionally called
forbidden beta transitions \footnote{%
Forbidden does not indicate that the transition is ompletely not allowed.
The vast contribution to $\beta $-decay transition is due to $s$-state
leptonic emission.}.

The general $\beta $-decay $N_{1}\rightarrow N_{2}+Lepton+antiLepton$ means
each lepton carries spin $s=\frac{1}{2}$ . The $\beta ^{\pm }$decay final
leptonic state can couple to total spin $s_{leptons}=0,1$. In EC process the
initial proton and electron can couple to $j\pm \frac{1}{2}$ and the final
neutrino can couple to $j\pm \frac{1}{2}$ \ or $j\mp \frac{1}{2}$. Thus in
all cases the lepton \ spin can change nuclear total angular moment $J$ by 0
or 1. In allowed $\beta $-decay, transitions with no angular momentum change
are called Fermi transitions and those with total angular momentum change by
one unit are called Gamow-Teller transitions \cite{schopper60}. There is no
source for parity changing allowed $\beta $-deay transition $(-1)^{\Delta
l}=+1$ \cite{schopper60}. In the standard model the lepton number is considered to be conserved
separately for each lepton flavour: electron $e$, muon $\mu $, and tau $\tau$.

\begin{table}[htb]
\caption{Electric charge $q$, baryon number $B$, lepton number $L$, and mass 
$m$ for fermions involved in the $\beta$-decay.}
\label{tab_beta_fermions}
\par
\begin{center}
\begin{tabular}{lcccr}
\hline
particle & $q$ & $B$ & $L$ & $m$ (MeV/$c^{2}$) \\ \hline
electron $(e^{-})$ & $-e$ & $0$ & $+1$ & $0.511$ \\ 
positron $(e^{+})$ & $+e$ & $0$ & $-1$ & $0.511$ \\ 
electron neutrino $(\nu _{e})$ & $0$ & $0$ & $+1$ & $0$ \\ 
electron antineutrino $(\bar{\nu}_{e})$ & $0$ & $0$ & $-1$ & $0$ \\ 
proton $(p)$ & $+e$ & $+1$ & $0$ & $938.3$ \\ 
neutron $(n)$ & $0$ & $+1$ & $0$ & $939.6$ \\ \hline
\end{tabular}%
\end{center}
\end{table}

\begin{table}[tbh]
\caption{Selection rule for allowed $\beta $-decay transitions.
Here $J_{i}(J_{f})$ is the angular momentum of the initial (final) nuclear
state and correspondingly for the parity $\pi$ \cite{Behrens1982}.}
\label{tab_beta_selection_rules}
\par
\begin{center}
\begin{tabular}{lll}
\hline
Type of tranaition & $\Delta J=\left\vert J_{f}-J_{i}\right\vert $ & $%
\pi_{i}\pi _{f}$ \\ \hline
Fermi & 0 & +1 \\ 
Gamow-Teller & 1$(J_{i}=0$ or $J_{f}=0)$ & +1 \\ 
Gamow-Teller & 0,1$(J_{i}>0,J_{f}>0)$ & +1 \\ \hline
\end{tabular}%
\end{center}
\end{table}

\subsection{Half-lives, reduced transition probabilities, and $ft$ values}

Half-life represented by $t_{\frac{1}{2}}$ is computed from transition
probability $\mathcal{T}_{fi}$, 
\begin{equation}
t_{\frac{1}{2}}=\frac{\ln 2}{\mathcal{T}_{fi}},  \label{eq65}
\end{equation}%
$\mathcal{T}_{fi}$ is calculated Fermi golden rule of time-dependent
perturbation theory \cite{HKHSnucphysa97} to get
\begin{equation}
t_{1/2}=\frac{\kappa}{f_0(B_F+B_{GT})},
\label{eq66}
\end{equation}
where $\kappa $ (kappa) is a constant \cite{Behrens1982}
\begin{equation}
\kappa =\frac{2\pi ^{3}\hbar ^{7}\ln 2}{m_{e}^{5}c^{4}G_{F}^{2}}=6147s,
\label{eq67}
\end{equation}%
$f_{0}$ is the lepton kinematics phase space integral, $B_{F}$ and $B_{GT}$
are the Fermi and Gamwo-Teller reduced transition probabilities needed to be
calculated, respectively. They can be broken up into factors \cite{Suhonen2017},
\begin{equation}
B_{F}=\frac{g_{v}^{2}}{2J_{i}+1}\left\vert \mathcal{M}_{F}\right\vert ^{2},
\label{eq69}
\end{equation}
and
\begin{equation}
B_{GT}=\frac{g_{A}^{2}}{2J_{i}+1}\left\vert \mathcal{M}_{GT}\right\vert ^{2},
\label{eq70}
\end{equation}%
where $J_{i}$ is the total angular momentum (nuclear spin) of the initial
nuclear state. $g_{V}$ and $g_{A}$ are coupling constants for vector current
and axial current, respectively \cite{Suhonen2019}. $\mathcal{M}_{F}$ and $%
\mathcal{M}_{GT}$ are the interaction amplitudes. The quantity $f_{0}t_{%
\frac{1}{2}}$(written as $ft$ value) represents the allowed $\beta $-decay
transition. It depends on nuclear structure, which contained in the reduced
matrix elements. In ref.\cite{HKHSnucphysa97} it has been called the reduced
half-life or comparative half-life. The vector coupling constant $g_{V}=1.0$%
, its value is determined by conserved current $j^{\mu }=\frac{1}{2}\bar{\psi%
}\gamma ^{\mu }\psi $\cite{Suhonen2017}.

The factor $g_{A}=1.25$, is the axial vector coupling constant of the weak
interaction determined by partially conserved axial vector current $%
j_{A}^{\mu }=\frac{1}{2}\bar{\psi}\gamma ^{\mu }\gamma ^{5}\psi $. All those
currents are calculated using the standard model. $g_{A}$ is affected by
many-nucleon correlation (pairing residual interaction) value reduced by
20-30\% is some times used \cite{Suhonen2017}. The presence of both vector
and axial vector coupling constant in $t_{\frac{1}{2}}$relation (\ref{eq66})
reflects the parity non-conserving nature of the weak interaction \cite%
{WuPhysRev57}. Vectors have parity properties, $\vec{V}(-\vec{r})=-\vec{V}(%
\vec{r})$ under space inversion. On the other hand, axial vector (pseudo
vector) $\vec{A}$ are invariant under space inversion,
\begin{equation*}
\vec{A}(-\vec{r})=+\vec{A}(\vec{r}).
\end{equation*}

For lepton current the violation of parity conservation is maximal, and the
weak interaction amplitude for the leptonic contribution contain the
combination $V-A$ in equal division. This holds in the quark level of the
hadrons \cite{halzen84}. The hadronic current 
\begin{equation}
j\propto V-(\frac{g_{A}}{g_{V}})A=V-(1.25A).  \label{eq71}
\end{equation}%
Thus the $V-A$ current is proportional to%
\begin{equation*}
V-A\propto \bar{\psi}\gamma ^{\mu }\left( 1-\gamma ^{5}\right) \psi .
\end{equation*}%
The minus sign is an indication of the left-handedness of the Leptons
involved in the weak interactions. Since $ft$ value is very large can be
suppressed by logarithm,
\begin{equation}
\log ft=\log _{10}(f_{0}t_{\frac{1}{2}}\left[ s\right] ).  \label{eq72}
\end{equation}

\subsection{Wigner-Eckart theorem}

Assume $\mathcal{T}$ $_{q}^{(k)}$ is spherical tensor operator (such as
angular momentum operators) acts on angular momentum basis $\left\vert
jm\right\rangle $. Transition amplitude due to tensor operator is given by 
\cite{Wigner1993, sakurai2020}
\begin{equation*}
\left\langle \xi _{f};j_{f}m_{f}\left\vert T_{q}^{(k)}\right\vert \xi
_{i};j_{i}m_{i}\right\rangle =\mathcal{M}\delta _{m_{f},m_{i}+q}
\end{equation*}%
$\mathcal{M}=0$ unless $m_{f}=m_{i}+q$. This is the Wigner-Eckart Theorem.
The matrix elements of tensor operators with respect to angular -momentum
eigenstates satisfy \cite{sakurai2020}
\begin{equation}
\left\langle \xi ^{\prime };j^{\prime }m^{\prime }\left\vert \mathcal{T}%
_{q}^{(k)}\right\vert \xi ;jm\right\rangle =\langle jm;kq\left\vert
jk;j^{\prime }m^{\prime }\right\rangle \frac{\left\langle \xi ^{\prime
}j^{\prime }\left\Vert \mathcal{T}^{(k)}\right\Vert \xi j\right\rangle }{%
\sqrt{2j+1}},  \label{eq79}
\end{equation}%
where the double-bar matrix element is independent of $m$ and $m^{\prime }$,
and $q$. Here $\xi $ this amplitude represents transition from $\left\vert
\xi ;jm\right\rangle $ to $\left\vert \xi ^{\prime };j^{\prime
}m^{\prime }\right\rangle $. Before we present a proof of this theorem,let
us look at its significance. First, we see that the matrix element is
written as the product of two factors. The first factor is a Clebsch-Gordan
coefficient for adding $j$ and $k$ \ to get $\ j^{\prime }$. It depends only
on the geometry-that is, on the way the system is oriented with respect to
the $z$-axis. There is no reference whatsoever to the particular nature of
the tensor operator. The second factor does depend on the dynamics; for
instance, $\xi $ may stand for the radial quantum number, and its evaluation
may involve, for example, evaluation of radial integrals. On the other hand,
it is completely independent of the magnetic quantum numbers $m$, $m^{\prime
}$, and $q$, which specify the orientation of the physical system. To
evaluate $\langle \xi ^{\prime },j^{\prime }m^{\prime }|\mathcal{T}%
_{q}^{(k)}|\xi ,jm\rangle $ with various combination of $m$, $m^{\prime}$%
, and $q^{\prime }$ it is sufficient to know just one of them: all others
can be related geometrically because they are proportional to Clebsch-Gordan
coefficients, which are known. The common proportionality factor is $\langle
\xi ^{\prime },j^{\prime }\left\Vert \mathcal{T}^{(k)}\right\Vert \xi
j\rangle $, which makes no reference whatsoever to the geometric features.
The selection rules for the tensor operator matrix element can be
immediately read off from the selection rules for adding angular momentum.
Indeed, from the requirement that the Clebsch-Gordan coefficient be
nonvanishing, we immediately obtain the $m$-selection rule derived before
and also the triangular relation $\left\vert j-k\right\vert \leq j^{\prime
}\leq j+k$.

There are different conventions for the reduced matrix elements. One
convention that includes an additional phase and normalization factor with
the aid of $6j$ symbol \cite{Wigner1993,Racah1942}%
\begin{equation}
\left\langle \xi ^{\prime };j^{\prime }m^{\prime }\left\vert \mathcal{T}%
_{q}^{(k)}\right\vert \xi ;jm\right\rangle =(-1)^{j-m}\left\{ 
\begin{array}{ccc}
j^{\prime } & k & j \\ 
-m^{\prime } & q & m%
\end{array}%
\right\} \left\langle \xi ^{\prime }j^{\prime }\left\Vert \mathcal{T}%
^{(k)}\right\Vert \xi j\right\rangle .  \label{q77}
\end{equation}

\subsection{Fermi and Gamow-Teller matrix element}

In the beginning let us review the scales of $\beta $-decay we need to
evaluate the transition matrices for. They are as follow
\begin{enumerate}
\item Quark scale. According to standard model the $\beta ^{-}$-decay is
attributed to weak flavor symmetry of $u$ and $d$ quarks, according to%
\begin{equation*}
u\rightarrow d+e+\bar{\nu}_{e}.
\end{equation*}

\item Nucleon scale. The $\beta ^{-}$-decay is due to decay of free (or
quasi-free) neutron, according to%
\begin{equation*}
n\rightarrow p+e+\nu _{e}.
\end{equation*}

\item Nuclear scale, where the $\beta ^{-}$-decay is due to the following
nuclear decay%
\begin{equation*}
_{_{Z}}^{^{A}}\text{X}\rightarrow _{Z+1}^{A}\text{Y}+e+\bar{\nu}_{e}.
\end{equation*}
\end{enumerate}

For nucleon scale $\beta ^{-}$-decay, we denote the proton using index $a$
or $f$, and the neutron using index $b$ or $i$. Whereas for $\beta ^{+}$%
-decay,we denote the neutron using index $a$ or $f$, and the proton using
index $b$ or $i$.

Fermi matrix element $\mathcal{M}_{F}$ \cite{fermi34} and Gamow-Teller (GT)
matrix element $\mathcal{M}_{GT}$ \cite{GTPhysRev36} are the most important
values needed to be calculated using the initial and final nuclear wave
functions which carry the nuclear structure information. Fermi operator is
just the unit operator $\hat{\mathbf{1}}$. GT operator is the Pauli spin
operator $\hat{\boldsymbol{\sigma}}$. These operators are the simplest scalar
and axial vector operators that can be constructed. The selection rules are
shown in table (\ref{tab_beta_selection_rules}).

The Fermi and Gamow-Teller matrix can be written as \cite{Suhonen2017}
\begin{equation}
\mathcal{M}_{F}=\left\langle \xi _{f}J_{f}\left\Vert \hat{\mathbf{1}}%
\right\Vert \xi _{i}J_{i}\right\rangle =\delta _{J_{i}J_{f}}\sum_{a,b}%
\mathcal{M}_{F}(fi)\left\langle \xi _{f}J_{f}\left\Vert \left[
c_{f}^{\dagger }\tilde{c}_{i}\right] _{\Delta J=0}\right\Vert \xi
_{i}J_{i}\right\rangle ,  \label{eq73}
\end{equation}
and 
\begin{equation}
\mathcal{M}_{GT}=\left\langle \xi _{f}J_{f}\left\Vert \hat{\boldsymbol{\sigma}}%
\right\Vert \xi _{i}J_{i}\right\rangle =\sum_{a,b}\mathcal{M}%
_{GT}(fi)\left\langle \xi _{f}J_{f}\left\Vert \left[ c_{f}^{\dagger }\tilde{c%
}_{i}\right] _{\Delta J=1}\right\Vert \xi _{i}J_{i}\right\rangle ,
\label{eq74}
\end{equation}%
where $\mathcal{M}_{F}(fi)$ and $\mathcal{M}_{GT}(fi)$ are the
single-particle matrix for Fermi and GT, respectively. They can be written
as \cite{Wigner1993,Racah1942}
\begin{eqnarray}
\mathcal{M}_{F}(ab) &=&\mathcal{M}_{F}(fi)=\left\langle f\left\Vert
\hat{\mathbf{1}}\right\Vert i\right\rangle =\delta _{fi}\hat{j}_{f}  \nonumber
\\
&=&\left\langle n_{f}l_{f}j_{f}\left\Vert \hat{\mathbf{1}}\right\Vert
n_{i}l_{i}j_{i}\right\rangle =\delta _{n_{f}n_{i}}\delta _{l_{f}l_{i}}\delta
_{j_{f}j_{i}}\hat{j}_{f},  \label{eq75}
\end{eqnarray}
and \cite{Wigner1993,Racah1942} 
\begin{eqnarray}
\mathcal{M}_{GT}(ab) &=&\mathcal{M}_{GT}(fi)=\frac{1}{\sqrt{3}}\left(
f\left\Vert \boldsymbol{\hat{\sigma}}\right\Vert i\right) =\frac{1}{\sqrt{3}}%
\left\langle n_{f}l_{f}j_{f}\left\Vert \boldsymbol{\hat{\sigma}}\right\Vert
n_{i}l_{i}j_{i}\right\rangle   \nonumber \\
&=&\frac{1}{\sqrt{3}}\sqrt{\frac{3}{2}}\times 2\delta _{n_{f}n_{i}}\delta
_{l_{f}l_{i}}\hat{\jmath}_{_{f}}\hat{\jmath}_{i}(-1)^{l_{f}+j_{f}+\frac{3}{2}%
}\left\{ 
\begin{array}{ccc}
\frac{1}{2} & \frac{1}{2} & 1 \\ 
j_{f} & j_{i} & l_{f}%
\end{array}%
\right\} ,  \nonumber \\
&=&\sqrt{2}\delta _{n_{f}n_{i}}\delta _{l_{f}l_{i}}\hat{\jmath}_{_{f}}\hat{%
\jmath}_{i}(-1)^{l_{f}+j_{f}+\frac{3}{2}}\left\{ 
\begin{array}{ccc}
\frac{1}{2} & \frac{1}{2} & 1 \\ 
j_{f} & j_{i} & l_{f}%
\end{array}%
\right\} .  \label{eq76}
\end{eqnarray}

\subsection{Symmetry properties of SP-matrix element}

Since we do not have orbital degrees of freedom, the symmetry properties for
Fermi single-particle matrix is%
\begin{equation}
\mathcal{M}_{F}(ab)=\mathcal{M}_{F}(ba),  \label{eq79-1}
\end{equation}%
which means that Fermi transitions for $\beta ^{+}$and $\beta ^{-}$are
similar. For GT transition, we have
\begin{equation}
\mathcal{M}_{GT}(ab)=(-1)^{j_{a}+j_{b}+1}\mathcal{M}(ba).  \label{eq79-2}
\end{equation}%
The Gamow-Teller single-particle matrix element for the lowest $l_{j}$
combinations are independent of $n$ as long as $\Delta n=0$ and thus they
obey the selection rule $\Delta l=0$. Table (\ref{tab_GT_SP_Matrix}) gives
the GT SP-matrix. These matrix element are calculated using a C++ function
based on eq.(\ref{eq76}).

\begin{table}[tbh]
\caption{Gamow-Teller single-particle matrix elements.}
\label{tab_GT_SP_Matrix}
\par
\begin{center}
\begin{tabular}{c|cccccccc}
\hline
$a/b$ & $s_{\frac{1}{2}}$ & $p_{\frac{3}{2}}$ & $p_{\frac{1}{2}}$ & $d_{%
\frac{5}{2}}$ & $d_{\frac{3}{2}}$ & $f_{\frac{7}{2}}$ & $f_{\frac{5}{2}}$ & $%
g_{\frac{9}{2}}$ \\ \hline
$s_{\frac{1}{2}}$ & $\sqrt{2}$ & $0$ & $0$ & $0$ & $0$ & $0$ & $0$ & $0$ \\ 
$p_{\frac{3}{2}}$ & $0$ & $2\frac{\sqrt{5}}{3}$ & $-\frac{3}{4}$ & $0$ & $0$
& $0$ & $0$ & $0$ \\ 
$p_{\frac{1}{2}}$ & $0$ & $\frac{4}{3}$ & $-\frac{\sqrt{2}}{3}$ & $0$ & $0$
& $0$ & $0$ & $0$ \\ 
$d_{\frac{5}{2}}$ & $0$ & $0$ & $0$ & $\sqrt{\frac{14}{5}}$ & $-\frac{4}{%
\sqrt{5}}$ & $0$ & $0$ & $0$ \\ 
$d_{\frac{3}{2}}$ & $0$ & $0$ & $0$ & $\frac{4}{\sqrt{5}}$ & $-\frac{2}{%
\sqrt{5}}$ & $0$ & $0$ & $0$ \\ 
$f_{\frac{7}{2}}$ & $0$ & $0$ & $0$ & $0$ & $0$ & $2\sqrt{\frac{6}{7}}$ & $-4%
\sqrt{\frac{2}{7}}$ & $0$ \\ 
$f_{\frac{5}{2}}$ & $0$ & $0$ & $0$ & $0$ & $0$ & $4\sqrt{\frac{2}{7}}$ & $-%
\sqrt{\frac{19}{7}}$ & $0$ \\ 
$g_{\frac{9}{2}}$ & $0$ & $0$ & $0$ & $0$ & $0$ & $0$ & $0$ & $\frac{1}{3}%
\sqrt{\frac{110}{3}}$ \\ \hline
\end{tabular}%
\end{center}
\end{table}

\subsection{Phase-space factors}

The half-life contains the integrated leptonic phase space which is called a
phase-space factor $f_{0}$. Some references call it Fermi integral. For $%
\beta ^{\pm }$-decay, the phase-space factor is \cite{Gove1971}
\begin{equation}
f_{0}^{(\pm )}=\int_{0}^{E_{0}}F_{0}(\mp Z_{f},\varepsilon )p\varepsilon
(E_{0}-\mathcal{E})^{2}d\varepsilon ,  \label{eq80}
\end{equation}%
$F_{0}$ is the Fermi function. $\mathcal{E}$ is the energy ratio given by
\begin{equation}
\mathcal{E}=\frac{E_{e}}{m_{e}c^{2}},  \label{eq81}
\end{equation}%
where $E_{e}$ is the total energy of the emitted electron or positron. $E_{0}
$ denotes the nuclear energy difference 
\begin{equation}
E_{0}=\frac{E_{i}-E_{f}}{m_{e}c^{2}},  \label{eq82}
\end{equation}%
where $E_{i}$ and $E_{f}$ are the initial and final energies, respectively,
for nuclear states. The momentum is given by
\begin{equation}
p=\sqrt{\mathcal{E}^{2}-1}.  \label{eq83}
\end{equation}

\begin{figure}[ptbh]
\begin{center}
\includegraphics[scale=1.0]{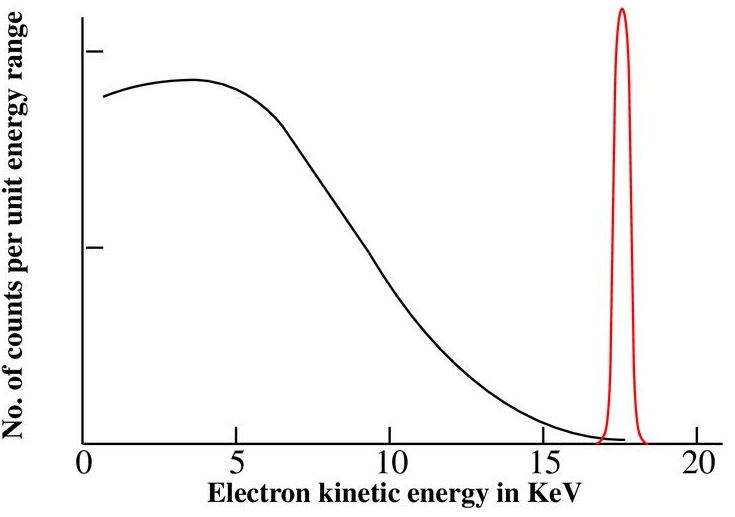}
\end{center}
\par\vspace{-0.7cm}
\caption{The $\beta $ decay
spectrum of tritium $(^{3}$H$\rightarrow ^{3}$He$)$. The red sharp spectra
is for nutrinoless decay. Taken from \cite{Lewis1970}.}
\label{fig_beta_spectrum}
\end{figure}

For electron capture the phase-space factor is \cite{Gove1971}%
\begin{equation}
f_{0}^{(EC)}=2\pi (\alpha Z_{i})^{3}(\mathcal{E}_{0}+E_{0})^{2},
\label{eq84}
\end{equation}%
where
\begin{equation}
\mathcal{E}_{0}=\frac{m_{e}c^{2}-\mathcal{B}}{m_{e}c^{2}}\approx 1-\frac{1}{2%
}(\alpha Z_{i})^{2},  \label{eq85}
\end{equation}%
where $\mathcal{B}$ is the atomic binding energy of the \ captured electron
usually ($1s$ orbital) and \ 
\begin{equation}
\alpha =\frac{e^{2}/4\pi \epsilon _{0}}{\hbar c}=\frac{1}{137},  \label{eq86}
\end{equation}%
is the fine structure constant. The approximation for $\mathcal{E}_{0}$
in eq.(\ref{eq85}) is valid for $\alpha Z_{i}\ll 1$, which holds for light
nuclei $Z_{i}<40$. The phase-space factors eq.(\ref{eq80}) or eq.(\ref{eq84}%
) are functions of the nuclear energy difference $E_{0}$. The final state
for $\beta ^{\pm }$-decay involves a three body state (one baryon and two
leptons), which reflects a complicated kinematics in $E_{0}$ dependence of $f_{0}^{(\pm )}$.
In EC the final state is a two-body state and the energy-momentum conservation
result in a definite energy for the emitted neutrino. The $\beta ^{\pm }$-decay
spectrum of the emitted electron or positron continuous due the distribution
of energy among three body system. An example of the spectrum is shown in
fig.(\ref{fig_beta_spectrum}).

The Fermi function (\ref{eq80}) can be approximated using non-relativistic
analytical technique, known as \emph{Primakoff-Rosen approximation} \cite%
{primakoff1959}
\begin{equation}
F_{0}(Z_{f},\mathcal{E})\approx \frac{\mathcal{E}}{p}F_{o}^{(PR)}(Z_{f});%
\qquad F_{0}^{(PR)(\pm )}(\mp Z_{f})=\frac{2\pi \alpha Z_{f}}{1-e^{\mp 2\pi
\alpha Z_{f}}}.  \label{eq87}
\end{equation}%
This approximation yields good results for high $Q$-vlaues. We can expand
the phase-space factor in eq.(\ref{eq80}) \cite{Gove1971,primakoff1959}%
\begin{equation}
f_{0}^{(\pm )}\approx \frac{1}{30}\left(
E_{0}^{5}-10E_{0}^{2}+15E_{0}-6\right) F_{0}^{(PR)(\pm )}(\mp Z_{f}).
\label{eq87_1}
\end{equation}

\subsection{$\beta $-decay $Q$-values}

The $Q$-value for any nuclear reaction or decay is given by%
\begin{equation}
Q=K_{f}+K_{i}=E_{i}-E_{f}.  \label{eq88}
\end{equation}%
Using eq.(\ref{eq82}), For $\beta ^{-}$-decay we have%
\begin{equation}
E_{0}=\frac{Q_{\beta ^{-}}+m_{e}c^{2}}{m_{e}c^{2}}.  \label{eq89}
\end{equation}%
For $\beta ^{+}$-decay we have

\begin{equation}
E_{0}=\frac{Q_{\beta ^{+}}+m_{e}c^{2}}{m_{e}c^{2}}.  \label{eq90}
\end{equation}%
Finally, for EC
\begin{equation}
E_{0}=\frac{Q_{Ec}-m_{e}c^{2}}{m_{e}c^{2}}.  \label{eq 91}
\end{equation}%
The $Q$-values for all decays are listed in \cite{Gove1971}. $E_{0}$ represents the end point energy of the
decay. The decay half-life can be calculated directly once the one-body transition densities 
\begin{equation}
\left\langle \xi _{f}J_{f}\left\Vert \left[ c_{a}^{\dagger }\tilde{c}_{b}%
\right] \right\Vert \xi _{i}J_{i}\right\rangle ,  \label{eq 92}
\end{equation}%
are known.

\subsection{Classification of $\beta $-decay}

The classifications of $\beta $-decay is done interms of $\log ft$ values,
given in table (\ref{tab_beta_class}) \cite{Wu1966}.

\subsubsection{Superallowed transitions}

Take place for light nuclei such as $_{1}^{3}$H$,^{14}$C$,^{15}$C$,\ldots $,
where all protons and final neutrons at Fermi level results an overlap in
the initial and final nuclear wave function. This means the that the
transitions are of the SP-type and yields maximum value of the F and GT
matrix element.

\subsubsection{$l$-forbidden allowed transitions}

This type occur in case where simple SP transition in mean-field shell-model
picture, forbidden by $\Delta l=0$ \ selection rule included in eq.(\ref%
{eq75}) and eq.(\ref{eq76}) The selection rules in Table (\ref%
{tab_beta_selection_rules}) are fulfilled. This means the forbiddingness is
due to a single configuration approximate for $\Psi _{i}$ and $\Psi _{f}$.
Using a configuration mixing based on the residual interaction, such as
pairing effect, removes this forbiddingness and gives a finite value for $\log ft$,
is usually below 5 due to the lack of strength in the
configuration mixing \cite{de-Shalit1963}.

\subsubsection{Unfavorable allowed transitions}

Such transitions do not belong to either of the two types discussed above.
They are allowed SP transitions in that there is no $l$ forbiddingness.
However, the SP transitions are suppressed in the $\Psi _{i}$ and $\Psi _{f}$
due to the residual interaction.

\begin{table}[tbh]
\caption{Classification of $\beta$-decay transitions.}
\label{tab_beta_class}
\par
\begin{center}
\begin{tabular}{lc}
\hline
Type of transition & $\log ft$ \\ \hline\hline
Superallowed & $2.9-3.7$ \\ 
Unfavoured allowed & $3.8-6.7$ \\ 
$l$-forbidden allowed & $\geq 5.0$ \\ 
First-forbidden unique & $8-10$ \\ 
First-forbidden non-unique & $6-9$ \\ 
Second-forbideen & $11-13$ \\ 
Third-forbidden & $17-19$ \\ 
Fourth-forbidden & $>22$ \\ \hline
\end{tabular}%
\end{center}
\end{table}

\section{Results and Discussion}

\subsection{$\beta -$Decay transition in one-particle and one-hole
nuclei}

This is the simplest possible nuclei, which considered stepping stone to
more complex structure.

\subsection{Matrix elements}

The wave functions of one-particle and one-hole \emph{nuclei} can be written as,
\begin{equation}
\left\vert \Psi _{i}\right\rangle =\left\vert
n_{i}l_{i}j_{i}m_{i}\right\rangle =c_{i}^{\dagger }\left\vert
CORE\right\rangle ,  \label{eq93}
\end{equation}%
and 
\begin{equation}
\left\vert \Psi _{f}\right\rangle =\left\vert
n_{f}l_{f}j_{f}m_{f}\right\rangle =c_{f}^{\dagger }\left\vert
CORE\right\rangle .  \label{eq94}
\end{equation}%
For the one-hole nuclei, the wave functions are,%
\begin{eqnarray}
\left\vert \Phi _{i}\right\rangle &=&\left\vert \left(
n_{i}l_{i}m_{i}\right) ^{-1}\right\rangle =h_{i}^{\dagger }\left\vert HF%
\right\rangle ,  \label{eq95} \\
\left\vert \Phi _{f}\right\rangle &=&\left\vert \left(
n_{f}l_{f}m_{f}\right) ^{-1}\right\rangle =h_{f}^{\dagger }\left\vert HF%
\right\rangle .  \notag
\end{eqnarray}

The one-body transition densities, derived according to Fermi golden rule are%
\begin{equation}
\left\langle \Psi _{f}\left\Vert \left[ c_{a}^{\dagger },\tilde{c}_{b}\right]
_{L}\right\Vert \Psi _{i}\right\rangle =\hat{L}\delta _{af}\delta _{bi},0
\label{eq95-1}
\end{equation}%
and%
\begin{equation}
\left\langle \Phi _{f}\left\Vert \left[ c_{a}^{\dag },\tilde{c}_{b}\right]
_{L}\right\Vert \Phi _{i}\right\rangle =\hat{L}\delta _{ai}\delta
_{bf}(-1)^{j_{i}+j_{f}+L}.  \label{eq95-2}
\end{equation}%
Here $\hat{L}$ is a Wigner-Eckart normalization 
\begin{equation}
\hat{L}=\frac{1}{\sqrt{2L+1}},  \label{eq95-3}
\end{equation}%
where $L$ is the resultant orbital angular momentum of the $L_{i}$ and $L_{f}
$ coupling. Recall that%
\begin{equation}
\left[ c_{a}^{\dagger },\tilde{c}_{b}\right] _{L}=\sum_{M_{\alpha },M_{\beta
}}\langle L_{a}M_{\alpha }L_{b}M_{\beta }|LM\rangle c_{\alpha }^{\dagger }%
\tilde{c}_{\beta }|CORE\rangle ,  \label{eqCG_commutator}
\end{equation}%
and using (\ref{eq73}) and (\ref{eq74}) we have%
\begin{equation}
\mathcal{M}_{F}=\left\langle \xi _{f}J_{f}\left\Vert \hat{1}\right\Vert \xi
_{i}J_{i}\right\rangle =\delta _{J_{i}J_{f}}\sum_{a,b}\mathcal{M}%
_{F}(ab)\left\langle \xi _{f}J_{f}\left\Vert \left[ c_{a}^{\dagger }\tilde{c}%
_{b}\right] _{\Delta J=0}\right\Vert \xi _{i}J_{i}\right\rangle ,
\label{eq95-4}
\end{equation}%
and 
\begin{equation}
\mathcal{M}_{GT}=\left\langle \xi _{f}J_{f}\left\Vert \hat{\sigma}%
\right\Vert \xi _{i}J_{i}\right\rangle =\sum_{a,b}\mathcal{M}%
_{GT}(ab)\left\langle \xi _{f}J_{f}\left\Vert \left[ c_{a}^{\dagger }\tilde{c%
}_{b}\right] _{\Delta J=1}\right\Vert \xi _{i}J_{i}\right\rangle .
\label{eq95-5}
\end{equation}

According to (\ref{eq79-1}) and (\ref{eq79-2}), there are symmetry relations
between one-particle and one-hole amplitude, 
\begin{equation}
\mathcal{M}_{F}(\Psi _{i}\rightarrow \Psi _{f})=-\mathcal{M}_{F}(\Phi
_{i}\rightarrow \Phi _{f})=\delta _{if}\hat{\jmath}_{i},
\label{eq_symmetryF}
\end{equation}%
and 
\begin{equation}
\mathcal{M}_{GT}(\Psi _{i}\rightarrow \Psi _{f})=\mathcal{M}_{GT}(\Phi
_{i}\rightarrow \Phi _{f})=\sqrt{3}\mathcal{M}_{GT}(ab),
\label{eq_symmetryGT}
\end{equation}%
where $\mathcal{M}_{F}$ and $\mathcal{M}\ _{GT}(ab)$ are single particle
(SP) Fermi and Gamow-Teller matrix elements. respectively. These are
substituted into eq.(\ref{eq69}) and eq.(\ref{eq70}), which yields%
\begin{equation}
B_{F}=\frac{g_{v}^{2}}{2J+1}\left\vert \mathcal{M}_{F}\right\vert ^{2},
\label{eq96}
\end{equation}%
and
\begin{equation}
B_{GT}=\frac{g_{A}^{2}}{2J+1}\left\vert \mathcal{M}_{GT}\right\vert ^{2},
\label{eq97}
\end{equation}%
where we obtain
\begin{equation}
B_{F}=g_{_{V}}^{^{2}}\delta _{if},  \label{eq98}
\end{equation}%
and
\begin{equation}
B_{GT}=g_{A}^{2}\frac{3}{2J_{i}+1}\left\vert \mathcal{M}_{GT}(ab)\right\vert
^{2}=g_{A}^{2}\frac{3}{2J_{i}+1}\left\vert \mathcal{M}_{GT}(fi)\right\vert
^{2}.  \label{eq99}
\end{equation}%
The values of are taken from references \cite{Suhonen2017,Suhonen2019}. The
relation are valid for transitions between one-particle states and for
transitions between one-hole states.

\subsection{Calculating EC half-life for $^{15}$O and $^{15}$N isobars}

\begin{figure}[ptbh]
\begin{center}
\includegraphics[scale=0.75]{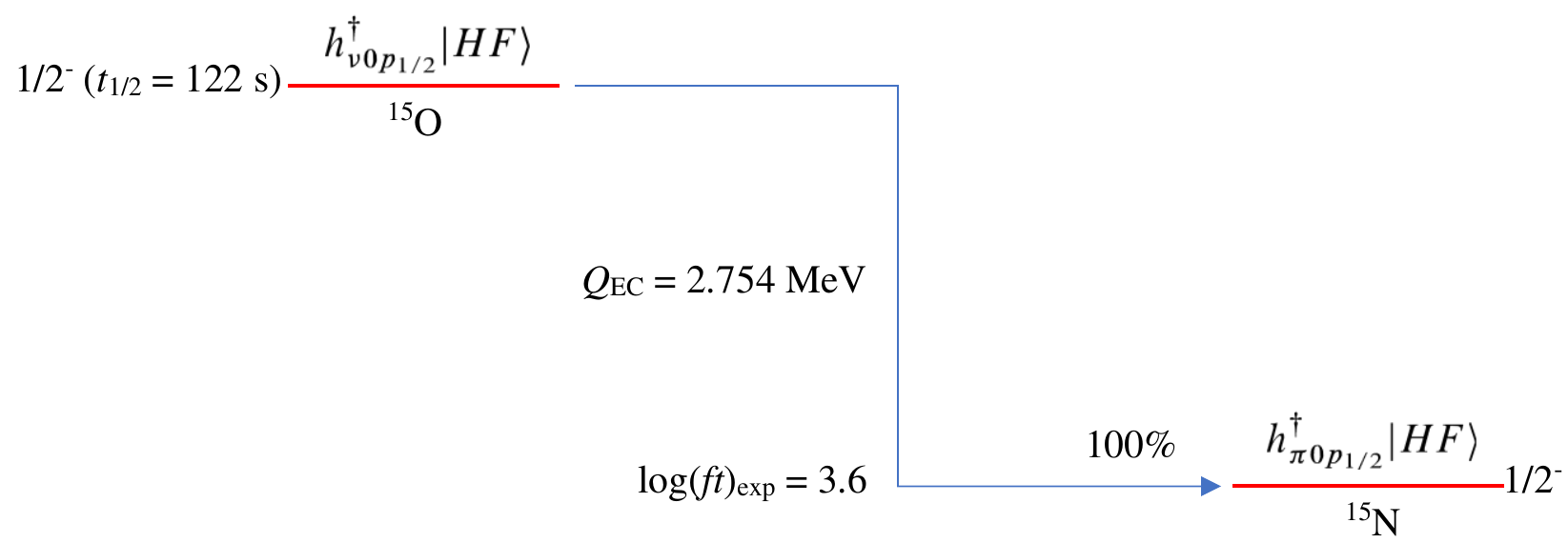}
\end{center}
\par\vspace{-0.7cm}
\caption{The decay scheme of $^{15}$O in
ground state to $^{15}$N ground state via the $\beta ^{+}/EC$ decay
mode. The experimental half-life, $Q$ value, branching and $\log (ft)$ value
are shown in the figure.}
\label{fig_O16_N15_EC}
\end{figure}

The electronic capture equation is%
\begin{equation*}
e+_{8}^{15}\text{O}_{7}\overset{EC}{\longrightarrow }_{7}^{15}\text{N}%
_{8}+\nu _{e}.
\end{equation*}%
We can calculate the $Q$-vlaue as%
\begin{equation*}
Q_{EC}=2.754\text{ MeV}.
\end{equation*}%
The experimental $\log ft$ value is 3.6 meaning the transition is
superallowed. The decay scheme and experimental information is shown in fig.(%
\ref{fig_O16_N15_EC}).

\begin{figure}[ptbh]
\begin{center}
\includegraphics[scale=0.75]{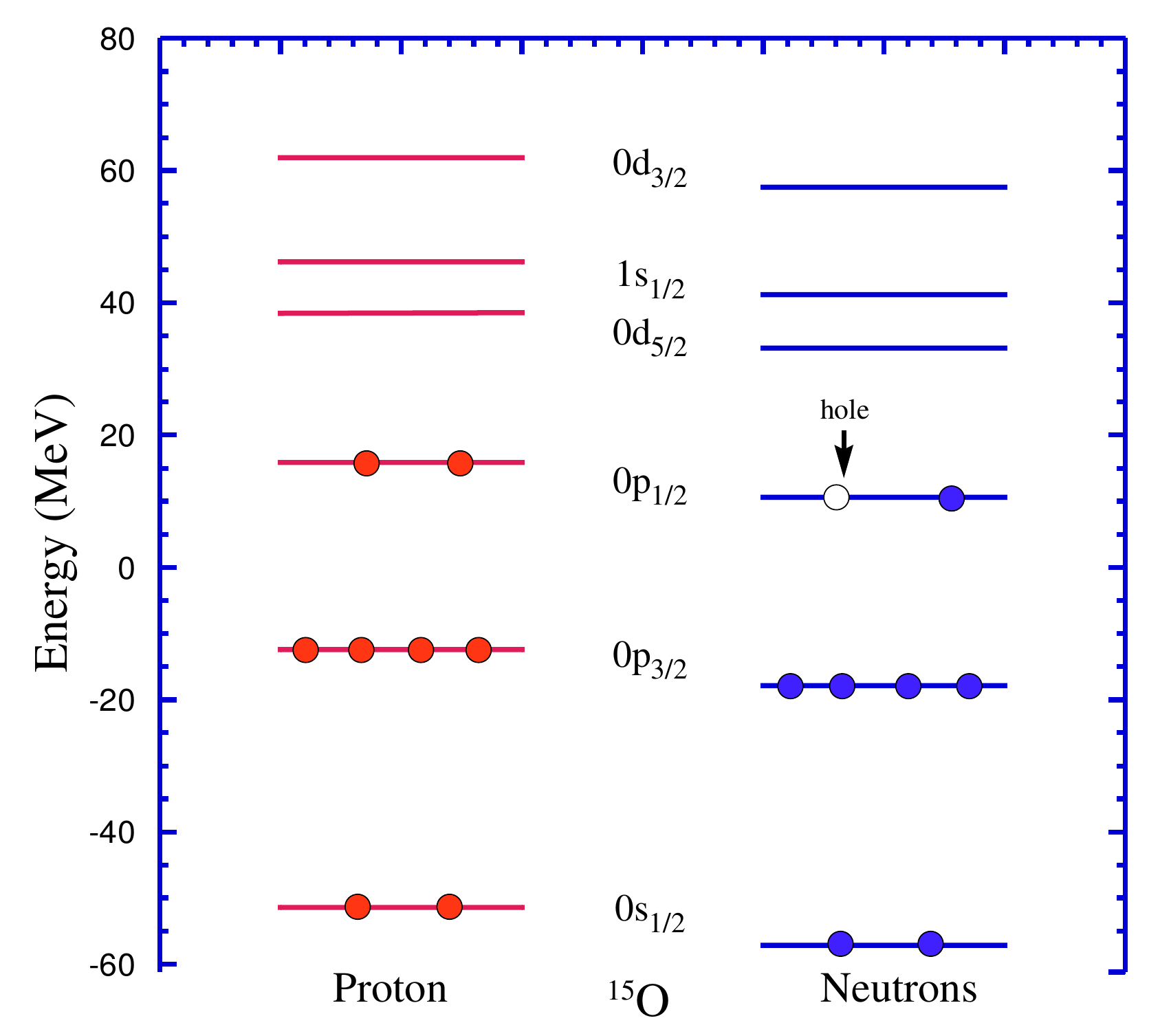}
\end{center}
\par\vspace{-0.7cm}
\caption{Single particle states of $^{15}$%
O generated using JISP6 potential in 8-shell model space. The SP particle
transition $\pi 0p_{1/2}\rightarrow \nu 0p_{1/2}$ energy is $%
\Delta E=5.221$ MeV.}
\label{fig_O15_SP}
\end{figure}

The Fermi SP-matrix is 
\begin{equation}
\mathcal{M}_{F}(ab)=\langle f\Vert \hat{1}\Vert i\rangle =\delta
_{fi}\hat{\jmath}_{a}=\left\langle n_{f}l_{f}j_{f}\left\Vert 1\right\Vert
n_{i}l_{i}j_{i}\right\rangle =\delta _{n_{f}n_{i}}\delta _{l_{f}l_{i}}\delta
_{j_{f}j_{i}}J_{b}.  \label{eq100}
\end{equation}
Using eq.(\ref{eq98}), the reduced Fermi matrix is thus 
\begin{equation}
B_{F}=g_{V}^{2}=1.0,  \label{eq100_1}
\end{equation}%
The GT SP-matrix is 
\begin{eqnarray}
\mathcal{M}_{GT}(fi) &=&\frac{1}{\sqrt{3}}\langle f\Vert \hat{%
\sigma}\Vert i\rangle =\frac{1}{\sqrt{3}}\left\langle
n_{f}l_{f}j_{f}\left\Vert \sigma \right\Vert n_{i}l_{i}j_{i}\right\rangle  
\notag \\
&=&\frac{1}{\sqrt{3}}\sqrt{\frac{3}{2}}\times 2\delta _{n_{f}n_{i}}\delta
_{l_{f}l_{i}}\hat{\jmath}_{f}\hat{\jmath}_{i}(-1)^{l_{f}+j_{i}+\frac{3}{2}%
}\left\{ 
\begin{array}{ccc}
\frac{1}{2} & \frac{1}{2} & 1 \\ 
j_{f} & j_{i} & L_{f}%
\end{array}%
\right\}\notag \\
&=&-\sqrt{\frac{2}{3}}.  \label{eq102}
\end{eqnarray}
The $6j$ symbol is calculated using Mathematica. Make use eq.(\ref{eq102})
into eq.(\ref{eq70}) the reduced GT transition matrix is thus%
\begin{equation}
B_{GT}=g_{A}^{2}\frac{3}{2}\left\vert \mathcal{M}_{GT}(\pi 0p_{\frac{1}{2}%
}\rightarrow \nu 0p_{\frac{1}{2}})\right\vert ^{2}=(1.25)^{2}\times \frac{3}{%
2}(-\sqrt{\frac{2}{3}})^{2}=0.521.  \label{eq101}
\end{equation}%
Make use eq.(\ref{eq100_1}) and eq.(\ref{eq101}) into eq.(\ref{eq66}), we
obtain
\begin{equation}
f_{0}t_{\frac{1}{2}}=\frac{6147s}{B_{F}+B_{GT}}=\frac{6147}{1.0+0.521}=4041.4%
\text{ s}.  \label{eq103}
\end{equation}%
Using the experimental half-life $t_{1/2}=122$ s, we can obtain the $f_{0}$
using eq.(\ref{eq66}) which yields
\begin{equation}
f_{0}=\frac{4041}{122}=33.12.  \label{eq104}
\end{equation}%
This corresponds to $\log (ft)$ value
\begin{equation}
\log (ft)=3.61,  \label{eq105}
\end{equation}%
as shown experimentally. The $f_{0}$ phase space function can be broken into
two phase space functions: one for $\beta ^{+}$-decay and the other for EC
decay. Therefore,
\begin{equation}
f_{0}=f_{0}^{(+)}+f_{0}^{EC}.  \label{eq106}
\end{equation}%
The nuclear energy difference in eq.(\ref{eq82}) is
\begin{equation}
E_{0}=\frac{Q_{EC^{-}}-m_{e}c^{2}}{m_{e}c^{2}}=\frac{2.754-0.511Mev}{0.511Mev%
}=4.389.  \label{eq107}
\end{equation}%
We can use Primakoff-Rosen approximation eq.(\ref{eq87}) \cite{primakoff1959}
\begin{equation}
F_{o}^{(PR)(\pm )}(\mp Z_{f})=\frac{2\pi \alpha Z_{f}}{1-e^{-2\pi \alpha
Z_{f}}}=\frac{2\pi \frac{1}{137}(7)}{1-e^{-2\pi \frac{1}{137}(7)}}=0.848,
\label{eq108}
\end{equation}%
with the aid of the phase space expansion (\ref{eq87_1}) we find $f_{0}^{(+)}$ 
\begin{equation}
f_{0}^{(+)}=42.3,  \label{eq109}
\end{equation}%
and use eq.(\ref{eq84}) we get%
\begin{equation}
f_{0}^{EC}=0.036.  \label{eq109-1}
\end{equation}%
We notice that $f$ $_{0}^{EC}<<$ \ $f_{0}^{(+)}$, which indicates that the
transition is dominated by $\beta ^{+}$. The total value of the phase factor
becomes%
\begin{equation}
f_{0}=f_{0}^{EC}+f_{0}^{(+)}=42.336,  \label{eq110}
\end{equation}%
Hence%
\begin{equation}
\log (f_{0})=1.63,  \label{eq110-1}
\end{equation}%
using the value of $\log (ft)$ in eq.(\ref{eq105}), we can calculate the
decay half-life%
\begin{equation}
t_{\frac{1}{2}}=10^{(\log ft-\log f_{0})}=10^{(.3.61-1.63)}=95.5\sec ,
\label{eq111}
\end{equation}%
This deviate from the experimental\ half-life ($t_{1/2}=122\sec $). The
previous steps for calculating the half-lives for one-particle-hole isotopes
are coded using C++ code named "beta\_decay\_halflife\_oneph.cpp" to
generate the data shown in table (\ref{tab_oph_halflife}).

\begin{table}[tbh]
\caption{Half-lives for odd-even (even-odd) isotopes computed using
one-particle-hole technique. The computed values are compared with the
experimental results.}
\label{tab_oph_halflife}
\par
\begin{center}
\begin{tabular}{|lccccc|}
\hline
Beta \ Decay & $Q_{EC}^{(\exp )}$(MeV) & $\log(f_{0})$ & $\log(ft)$ & $t_{%
\frac{1}{2}}$(s) & $t_{\frac{1}{2}}^{(\exp )}$(s) \\ \hline
$^{15}$O $(1/2^{+})\rightarrow $ $^{15}$N $(1/2^{-})$ & $2.754$ & $1.626$ & $%
3.606$ & $95.5$ & $122$ \\ 
$^{17}$F $(5/2^{+})\rightarrow $ $^{17}$O $(5/2^{+})$ & $2.762$ & $1.624$ & $%
3.283$ & $45.6$ & $64.5$ \\ 
$^{39}$Ca $(3/2^{+})\rightarrow $ $^{39}$K $(3/2^{+})$ & $6.524$ & $3.671$ & 
$3.500$ & $0.675$ & $0.86$ \\ 
$^{41}$Sc $(7/2^{-})\rightarrow $ $^{41}$Ca $(7/2^{-})$ & $6.495$ & $6.495$
& $3.308$ & $0.456$ & $0.59$ \\ \hline
\end{tabular}%
\end{center}
\end{table}

\subsection{Beta Decay to and from the Even-Even Ground state}

Charge-changing excitations of particle-hole nuclei can undergo beta decay
to the reference nucleus. The initial state is an odd-odd nucleus,
generated by making a charge-changing particle hole excitations of the
particle hole vacuum. Let the final state be the particle-hole vacuum
$\vert HF\rangle$, is the ground state for reference nucleus. Particle-hole
excited states are created by letting one nucleon jump from state below Fermi level to a state
above it \cite{lawson1980}. The wave functions of particle-hole nuclei is%
\begin{equation}
\left\vert ab^{-1};JM\right\rangle =\left[ c_{a}^{\dagger }h_{b}^{\dagger }%
\right] _{JM}\left\vert HF\right\rangle =\left[ c_{a}^{\dagger }\tilde{c}_{b}%
\right] _{JM}\left\vert HF\right\rangle ,  \label{eq55}
\end{equation}%
The wave functions is normalized%
\begin{equation}
\left\langle ab^{-1};JM|cd^{-1};J^{\prime }M^{\prime }\right\rangle =\delta
_{ac}\delta _{bd}\delta _{JJ^{\prime }}\delta _{MM^{\prime }}.  \label{eq56}
\end{equation}
Using (\ref{eq79}), with the aid of eq.(\ref{eq55}) and normalization (\ref{eq56}), The $\beta $%
-decay matrix elements are constructed from the transition density 
\begin{eqnarray}
\left\langle HF\left\Vert \left[ c_{a}^{\dagger }\tilde{c}_{b}\right]
_{L}\right\Vert a_{i}b_{i}^{-1};J_{i}\right\rangle  &=&\frac{\delta
_{LJ_{i}}\delta _{ab_{i}}\delta _{ba_{i}}(-1)_{i}^{j_{a_{i}}-j_{b_{i}}+J_{i}}%
}{\sqrt{2J_{i}+1}}  \notag \\
&=&\delta _{LJ_{i}}\delta _{ab_{i}}\delta
_{ba_{i}}(-1)_{i}^{j_{a_{i}}-j_{b_{i}}+J_{i}}\hat{J}_{i}  \label{eq112}
\end{eqnarray}%
Inserting eq.(\ref{eq95-4}) and eq.(\ref{eq95-5}), yields%
\begin{equation}
\mathcal{M}_{F}(a_{i}b_{i}^{-1})=\delta _{J_{i}0}\delta _{a_{i}b_{i}}\hat{%
\jmath}_{a_{i}},  \label{eq113}
\end{equation}%
and%
\begin{equation}
\mathcal{M}_{GT}(a_{i}b_{i}^{-1})=-\sqrt{3}\delta _{J_{i}1}\mathcal{M}%
_{GT}(a_{i}b_{i}),  \label{eq114}
\end{equation}%
where the symmetry relations (\ref{eq79-1}) and (\ref{eq79-2}) are used.
In case of the odd-odd nucleus has a low-lying states below the particle-hole
vacuum of the reference nucleus, $\beta $-decay occur from vacuum to the
odd-odd nucleus. This is common in light nuclei. Therefore eq.(\ref{eq112}) is
now replaced by%
\begin{equation}
\left\langle a_{f}b_{f}^{-1};J_{i}\left\Vert \left[ c_{a}^{\dagger }\tilde{c}%
_{b}\right] _{L}\right\Vert HF\right\rangle =\delta _{LJ_{f}}\delta
_{aa_{f}}\delta _{bb_{f}}\hat{J}_{f}.  \label{eq115}
\end{equation}%
Substituting eq.(\ref{eq95-4}) and eq.(\ref{eq95-5}), yields%
\begin{equation}
\mathcal{M}_{F}(a_{f}b_{f}^{-1})=\delta _{J_{f0}}\delta _{a_{f}b_{f}}~\hat{%
\jmath}_{a_{f}},  \label{eq116}
\end{equation}%
and%
\begin{equation}
\mathcal{M}_{GT}(a_{f}b_{f}^{-1})=-\sqrt{3}\delta _{J_{f1}}\mathcal{M}%
_{GT}(a_{f}b_{f}).  \label{eq117}
\end{equation}

\subsection{ Calculating the strength function of $^{56}$Ni decay}

\begin{figure}[ptbh]
\begin{center}
\includegraphics[scale=0.75]{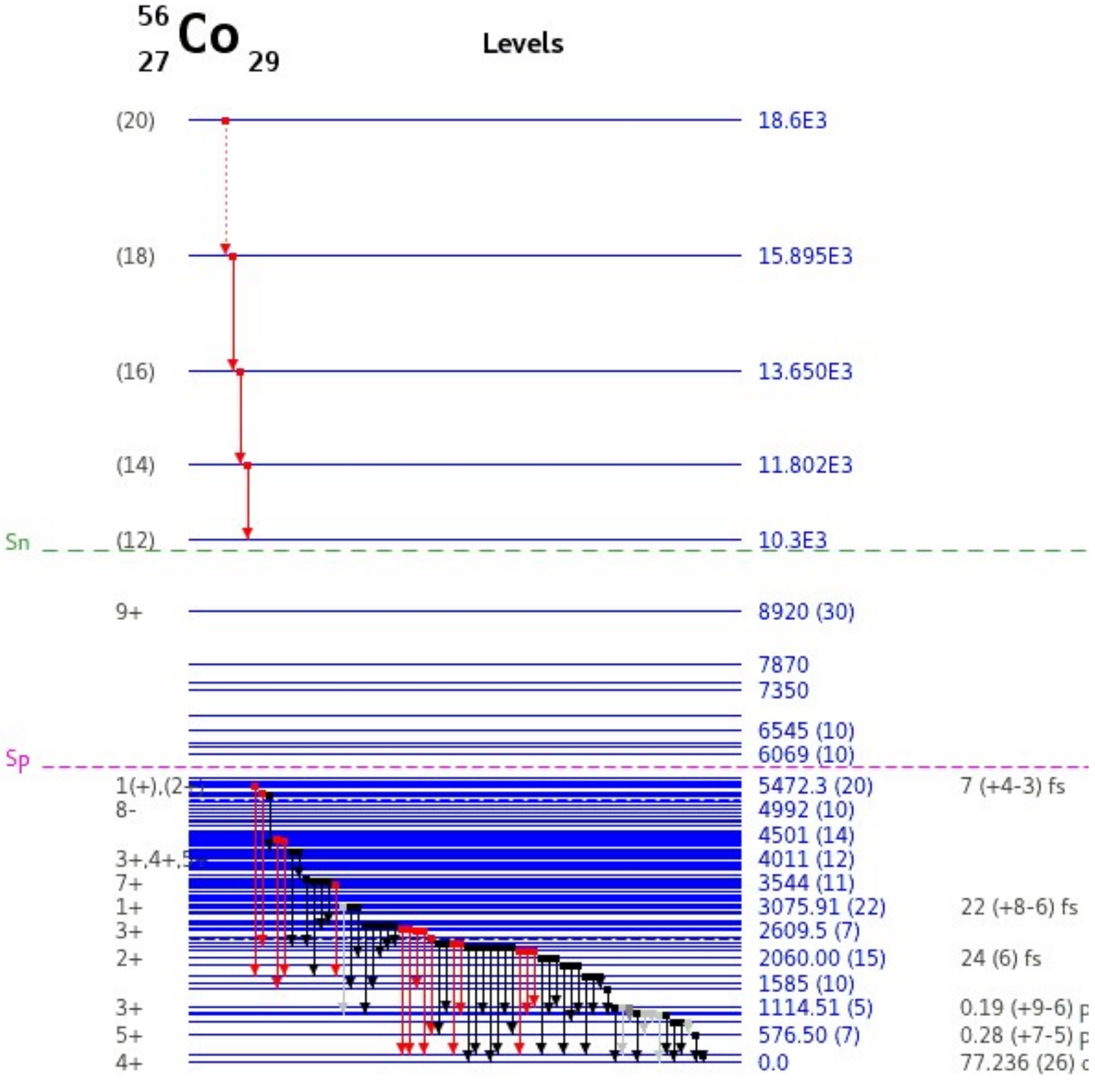}
\end{center}
\par\vspace{-0.7cm}
\caption{Energy levels of $^{56}$Co. The 
$^{56}$Co is the by-products of the $^{56}$Ni EC decay with $Q_{EC}=2.13$
MeV. Thus $^{56}$Co is formed in states with energy less than $E_{x}\leq 2.06
$ MeV. Taken from International Atomic Energy Agency.}
\label{fig_co56_states}
\end{figure}

The $\beta ^{+}/$ EC of
\begin{equation}
_{28}^{56}\text{Ni}_{28}+_{-1}^{0}e\rightarrow _{27}^{56}\text{Co}_{29}+\nu .
\label{eq_ni56_ec}
\end{equation}%
The $^{56}$Ni is in the ground state $0^{+}$, whereas $^{56}$Co can be
formed at excited states. The excited states of $^{56}$Co is shown in
fig.(\ref{fig_co56_states}). The $Q$-value of the decay is $Q_{EC}=2.13$ MeV.
This means the possible excited states $^{56}$Co can form at is 2.06 MeV.
This includes states $1^{+}$, $2^{+}$, $3^{+}$, $4^{+}$, $5^{+}$, and $6^{+}$.

\begin{figure}[ptbh]
\begin{center}
\includegraphics[scale=1.0]{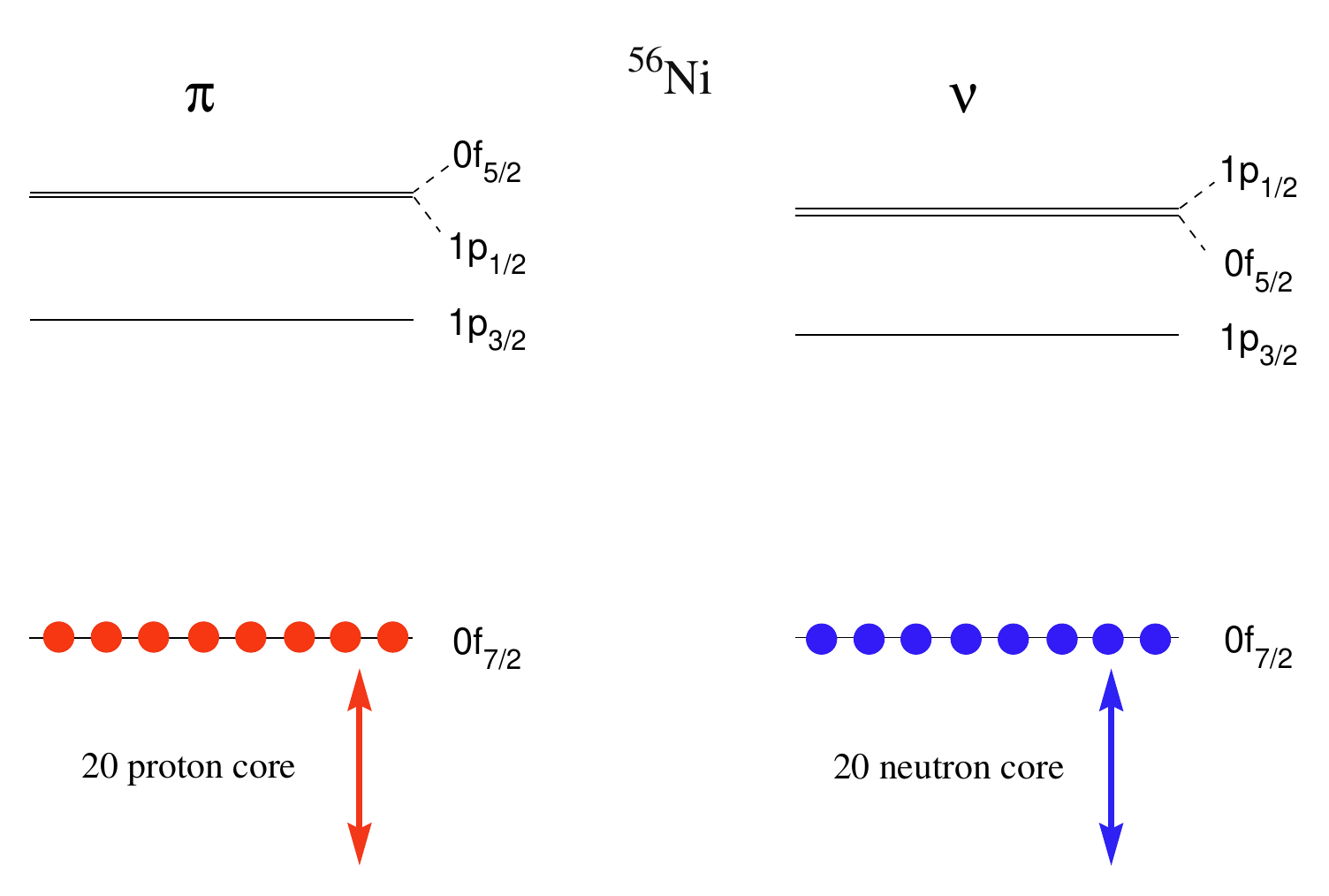}
\end{center}
\par\vspace{-0.7cm}
\caption{Valence shells $|HF\rangle $ state for $^{56}$Ni. At
ground state only $0f_{7/2}$ states are completely full with protons and
neutrons. The state spacings are proportional to the energy spacings
generated by Wood-Saxon potential in reference \cite{RHJphysrevc98}.}
\label{fig_Ni56_valence}
\end{figure}

In the SP scheme the valence of $^{56}$Ni in the ground state is shown in the
fig.(\ref{fig_Ni56_valence}). It shows that the core $|CORE\rangle _{\pi }$ is filled
with 20 protons and $|CORE\rangle _{\nu }$ is filled with 20 neutrons. The
valence states $|HF\rangle _{\pi }$ and $|HF\rangle _{\nu }$ has the $%
0f_{7/2}$ state full with 8 protons and 8 neutrons.

When one proton in the $\pi 0f_{7/2}$ state of $^{56}$Ni captures an
electron this causes vacancy in the $\pi 0f_{7/2}$ state of the daughter $%
^{56}$Co and an extra neutron is formed. Only this transition%
\begin{equation*}
\pi 0f_{7/2}\rightarrow \nu 0f_{5/2},
\end{equation*}%
is allowed by the SP transition matrix which set the selection rule for
either Fermi or GT transition, given in eq.(\ref{eq75}) and (\ref{eq76}).
Thus, the extra neutron must be formed in the $\nu 0f_{5/2}$ state of $^{56}$%
Co. In another word, a proton-hole created in the $\pi 0f_{7/2}$ and a
neutron-particle is created in the $\nu 0f_{5/2}$. This is shown in fig.(\ref%
{fig_Co56_byprod_valence}).

\begin{figure}[ptbh]
\begin{center}
\includegraphics[scale=1.0]{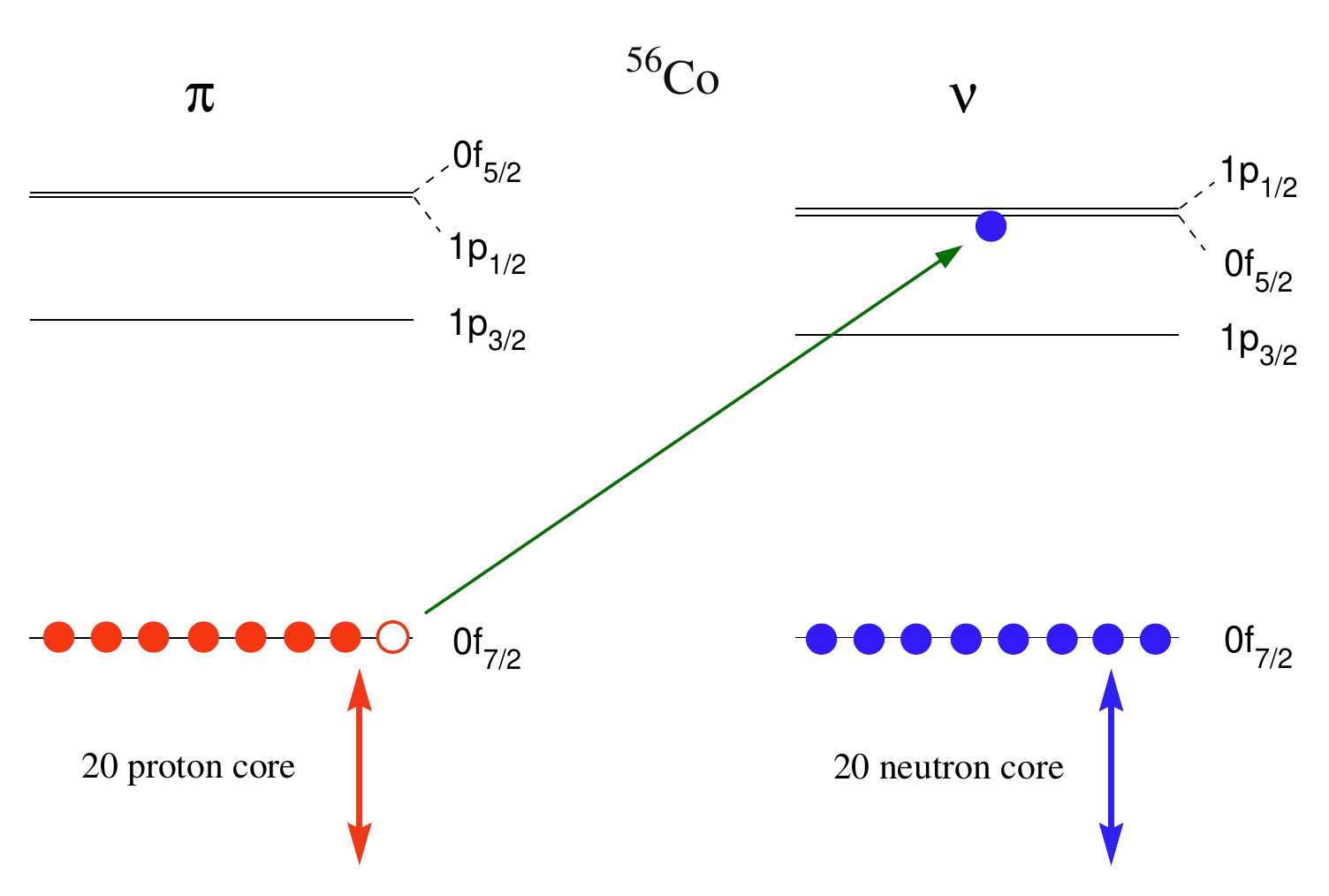}
\end{center}
\par\vspace{-0.7cm}
\caption{Valence shells $|HF\rangle $
state for $^{56}$Co. A proton hole $p^{-1}$ is created in the $\pi %
0f_{7/2}$ state ($h_{\pi 0f_{7/2}}^{\dagger }$) and a
neutron-particle is created in the in the $\nu 0f_{5/2}$ state ($c_{%
\nu 0f_{5/2}}^{\dagger }$). The state spacings are proportional to
the energy spacings generated by Wood-Saxon potential in reference 
\cite{RHJphysrevc98}.}
\label{fig_Co56_byprod_valence}
\end{figure}

The \emph{triangular condition }$\Delta ($ $J\frac{5}{2}\frac{7}{2})$ gives
all possible nuclear spin states of $^{56}$Co, thus%
\begin{equation*}
\left\vert \frac{7}{2}-\frac{5}{2}\right\vert =1\leq J\leq \left\vert \frac{7%
}{2}+\frac{5}{2}\right\vert =6.
\end{equation*}%
Henceforth, the nuclear state of $^{56}$Co is defined in terms of the ground
state of the nuclear state of $^{56}$Ni as%
\begin{equation}
\left\vert ^{56}\text{Co,}1^{+},2^{+},3^{+},4^{+},6^{+}\right\rangle =\left[
c_{\nu 0f_{\frac{5}{2}}}^{\dagger }h_{\pi 0f_{\frac{7}{2}}}^{\dagger }\right]
_{1^{+},2^{+},3^{+},4^{+},5^{+},6^{+}}\left\vert ^{56}\text{Ni}%
;0^{+}\right\rangle .  \label{eq_co56_ni56}
\end{equation}%
Since $\delta _{J_{f}0}=0\Rightarrow \mathcal{M}_{F}=0$ in eq.(\ref{eq116}),
only possible final state is for Gamow-Teller matrix is the $1^{+}$ state of 
$^{56}$Co according to eq.(\ref{eq117}). Thus the GT SP transition matrix
according to eq.(\ref{eq117}) is
\begin{equation}
\mathcal{M}_{GT}=\sqrt{3}\mathcal{M}_{GT}(f_{\frac{5}{2}}f_{\frac{7}{2}})=-%
\sqrt{3}(-4\sqrt{\frac{2}{7}})=3.703,  \label{eq118}
\end{equation}%
thus,%
\begin{equation}
B_{F}=0.  \label{eq119}
\end{equation}%
Make use of eq.(\ref{eq99}) into eq.(\ref{eq118}) the reduced GT transition
matrix is thus, 
\begin{equation}
B_{GT}=g_{A}^{2}\frac{3}{2J_{i}+1}\left\vert \mathcal{M}_{GT}(%
fi)\right\vert ^{2}=(1.25)^{2}\frac{3}{(2\times 1)+1}(3.703)^{2}=21.43.
\label{eq120}
\end{equation}%
Make use of eq.(\ref{eq119}) and eq.(\ref{eq120}) into eq.(\ref{eq66}), we
obtain%
\begin{equation}
f_{0}t_{\frac{1}{2}}=\frac{\kappa }{(B_{F}+B_{GT})}=\frac{6147}{21.43}%
=286.84.  \label{eq121}
\end{equation}%
Using eq.(\ref{eq72}) and eq.(\ref{eq121}), we obtain, 
\begin{equation}
\log ft=\log 286.84=2.46,  \label{eq122}
\end{equation}%
The experimental value, 
\begin{equation}
(\log ft)_{\exp }=4.4.  \label{eq123}
\end{equation}%
This is much less than the experimental value $\log ft=4.4$. An indication
for the failure of the particle-hole theory. According to table (\ref%
{tab_beta_class}), the transition is unfavoured allowed, meaning that the
single-particle transition is suppressed during the initial and the final
states due to the residual two-body interaction by a factor $%
10^{2.46-4.4}=0.011$.

The coupling in the excitation%
\begin{equation*}
\left\vert ^{56}\text{Co };1^{+}\right\rangle =\left[ c_{\nu of_{\frac{5}{2}%
}}^{\dagger }h_{\pi of_{\frac{7}{2}}}^{\dagger }\right] \left\vert ^{56}%
\text{Ni };0^{+}\right\rangle ,
\end{equation*}%
is the only way to produce $1^{+}$state by exciting a proton from $0f_{\frac{%
7}{2}}$ shell to the $0f_{\frac{5}{2}}$ for neutrons. Therefore, The
discrepancy suggests the need for more better configurations consists of
two-particle-hole excitations which plays an active part in the low laying
states of $\ ^{56}$Co.

\subsection{$\beta $-decay transitions between two particle-hole
states}

This configuration is based on the electromagnetic transitions between two
arbitrary particle-hole states:%
\begin{equation*}
\left\vert a_{i}b_{i}^{-1};J_{i}\right\rangle \rightarrow \left\vert
a_{f}b_{f}^{-1};J_{f}\right\rangle ,
\end{equation*}%
due to an operator $\hat{O}_{L}$ in which the transition amplitude is
\cite{Rodin2002,Yoshida2003}%
\begin{eqnarray}
\left\langle a_{f}b_{f}^{-1};J_{f}\left\vert \hat{O}_{L}\right\vert
a_{i}b_{i}^{-1};J_{i}\right\rangle  &=&(-1)^{j_{a_{i}}+j_{b_{f}}}\hat{J}_{i}%
\hat{J}_{f}\times   \nonumber \\
&&\left[ 
\begin{array}{c}
\delta _{b_{i}b_{f}}(-1)^{J_{i}+L}\left\{ 
\begin{array}{ccc}
J_{i} & J_{f} & L \\ 
j_{a_{f}} & j_{a_{i}} & j_{b_{i}}%
\end{array}%
\right\} \left\langle a_{i}\left\vert \hat{O}_{L}\right\vert
a_{f}\right\rangle + \\ 
\delta _{a_{i}a_{f}}(-1)^{J_{f}+L}\left\{ 
\begin{array}{ccc}
J_{i} & J_{f} & L \\ 
j_{b_{f}} & j_{b_{i}} & j_{a_{i}}%
\end{array}%
\right\} \left\langle b_{i}\left\vert \hat{O}_{L}\right\vert
b_{f}\right\rangle 
\end{array}%
\right] .  \label{eq_tph_em}
\end{eqnarray}
Here the $a$'s and the $b$'s are replaced by $\pi $'s and $\nu $'s.

Starting from even-even reference nucleus $(N-Z)$, has a particle hole
vacuum state. The excited states are proton-particle-hole $(pp^{-1})$ and
neutron-particle-hole $(nn^{-1})$ excitations. Consider $\beta ^{-}$decay of
the adjacent odd-odd $(Z+1,$ $N-1)$ we have $(np^{-1})$ excitations. Let as
define $\beta ^{-}$decay operator $\hat{\beta}_{LM}^{-}$ as 
\begin{equation}
\hat{\beta}_{LM}^{-}=(\hat{L})^{-1}\sum_{pn}\left\langle
p\right\Vert \hat{\beta}_{L}\left\Vert n\right\rangle \left[ c_{p}^{\dagger }%
\tilde{c}_{n}\right] _{LM}  \label{eq124}
\end{equation}%
for $L=0$ is Fermi operator and for $L=1$ Gamow-Teller operator. Thus $\beta
_{0}=\mathbf{1}$ and $\beta _{1}=\boldsymbol{\sigma }$. The transition
amplitude for state $\left\vert \Psi _{i}\right\rangle $ to $\left\vert \Psi
_{f}\right\rangle $ is given by 
\begin{equation}
\left\langle \Psi _{f}\right\Vert \beta _{L}^{-}\left\Vert \Psi
_{i}\right\rangle =(\hat{L})^{-1}\sum_{pn}\left\langle p\right\Vert
\beta _{L}\left\Vert n\right\rangle \left\langle \Psi
_{f}\right\Vert \left[ c_{p}^{\dagger }\tilde{c}_{n}\right] \left\Vert \Psi
_{i}\right\rangle .  \label{eq125}
\end{equation}%
We have the following different $\beta $-decay cases

\subsubsection{The initial state is a neutron-particle-proton-hole ($n_{i}p_{i}^{-1}$)}

The initial nuclear wave function is%
\begin{equation}
\left\vert \Psi _{i}\right\rangle =\left[ c_{n_{i}}^{\dagger
}h_{p_{i}}^{\dagger }\right] _{J_{i}M_{i}}\left\vert HF\right\rangle .
\label{eq125-1}
\end{equation}%
For the final nuclear wave function we need to consider two sub cases:
either the final state be neutron-particle-neutron-hole ($n_{f}n_{f}^{\prime
-1}$) or proton-particle-proton-hole ($p_{f}p_{f}^{\prime -1}$).

\paragraph{Neutron-particle-neutron-hole final state ($n_{f}n_{f}^{\prime-1} $)}
Here the final nuclear wave functions is
\begin{equation}
\left\vert \Psi _{f}\right\rangle =\left[ c_{n_{f}}^{\dagger
}h_{n_{f}^{\prime }}^{\dagger }\right] _{J_{f}M_{f}}\left\vert
HF\right\rangle .  \label{eq125-2}
\end{equation}%
Make use into ( \ref{eq125-1}) (\ref{eq125-2}) into ( \ref{eq125} ) use
Wigner-Eckart theorem 
\begin{eqnarray}
\left\langle n_{f}n_{f}^{^{\prime }-1};J_{f}\left\Vert \beta
_{L}^{-}\right\Vert n_{i}p_{i}^{-1};J_{i}\right\rangle  &=&\delta
_{n_{f}n_{i}}(-1)^{j_{n_{i}}+j_{n_{f}^{\prime }}+J_{f}+1}\hat{J}_{i}\hat{J}%
_{f}\hat{L}\times   \nonumber \\
&&\left\{ 
\begin{array}{ccc}
J_{i} & J_{f} & L \\ 
j_{n_{f}^{\prime }} & j_{p_{i}} & j_{n_{i}}%
\end{array}%
\right\} \mathcal{M}_{L}(p_{i}n_{f}^{\prime }).\label{eq126}
\end{eqnarray}
Note that $n_{f}^{\prime }$ \ represents neutron hole $n^{-1}$.
$\mathcal{M}_{L}($ $p_{i}n_{f}^{^{\prime }}$ $)$ is either the Fermi $(L=0)$
or the Gamow-Teller $(L=1)$ single-particle matrix element given in \ref%
{eq73} and \ref{eq74}, respectively.
\begin{equation}
\mathcal{M}_{0}(ab)=\mathcal{M}_{F}(ab)=\left\langle a\left\Vert \beta
_{0}\right\Vert b\right\rangle =\left\langle a\left\Vert 1\right\Vert
b\right\rangle ,  \label{eq127}
\end{equation}%
\begin{equation}
\mathcal{M}_{1}(ab)=\mathcal{M}_{GT}(ab)=\frac{1}{\sqrt{3}}\left\langle
a\left\Vert \beta _{1}\right\Vert b\right\rangle =\frac{1}{\sqrt{3}}%
\left\langle a\left\Vert \boldsymbol{\sigma }\right\Vert b\right\rangle .
\label{eq128}
\end{equation}%
In the single particle matrix element $\mathcal{M}_{L}$ proton and neutrons
labels are no longer distinct.

\paragraph{Proton-particle-proton-hole final state ($p_{f}p_{f}^{\prime -1}$)}

The final wave function
\begin{equation}
\left\vert \Psi _{f}\right\rangle =\left[ c_{p_{f}}^{\dagger
}h_{p_{f}^{\prime }}^{\dagger }\right] _{J_{f}M_{f}}\left\vert %
HF\right\rangle .  \label{eq129}
\end{equation}%
Make use into eq (\ref{eq125}), we obtain 
\begin{eqnarray}
\left\langle p_{_{f}}p_{f}^{\prime -1};J_{f}\left\Vert \beta
_{L}^{-}\right\Vert n_{i}p_{i}^{-1};J_{i}\right\rangle  &=&\delta
_{p_{i}p_{f}^{\prime }}(-1)^{j_{n_{i}}+j_{p_{f}^{\prime }}+J_{i}+L}\hat{J}%
_{i}\hat{J}_{f}\hat{L}\times   \nonumber \\
&&\left\{ 
\begin{array}{ccc}
J_{i} & J_{f} & L \\ 
j_{p_{f}} & j_{n_{i}} & j_{p_{i}}%
\end{array}%
\right\} \mathcal{M}_{L}(n_{i}p_{f}).  \label{eq130}
\end{eqnarray}

For $\beta ^{+}$decay, initial state odd-odd $(Z+1,N-1)$ nucleus generated
by proton- particle-neutron-hole $(pn^{-1})$ excitation of $(N,Z)$
particle-hole vacuum. The final state is particle-hole excitation, obeys
charge conservation condition, for the even-even reference nucleus $(Z,N)$.
In similar argument to eq (\ref{eq124}), the decay operator is:
\begin{equation}
\beta _{LM}^{+-}=\sum_{pn}\left\langle n\right\Vert \beta
_{L}\left\Vert p\right\rangle \left[ c_{n}^{\dagger }\tilde{c}_{p}\right]
_{LM}.  \label{eq130-1}
\end{equation}%
Thus the transition amplitude is similar to eq.(\ref{eq125}), which reads as
\begin{equation}
\left\langle \Psi _{f}\right\Vert \beta _{L}^{+-}\left\Vert \Psi
_{i}\right\rangle =(\hat{L})^{-1}\sum_{pn}\left\langle n\right\Vert
\beta _{L}\left\Vert p\right\rangle \left\langle \Psi _{f}\right\Vert \left[
c_{n}^{\dagger }\tilde{c}_{p}\right] \left\Vert \Psi _{i}\right\rangle .
\label{eq131}
\end{equation}

\subsubsection{The initial state is a proton-particle-neutron-hole state ($p_{i}n_{i}^{-1}$)}

In this case the initial nuclear wave function is
\begin{equation}
\left\vert \Psi _{i}\right\rangle =\left[ c_{p_{i}}^{\dagger
}h_{n_{i}}^{\dagger }\right] _{J_{i}M_{i}}\left\vert HF\right\rangle .
\label{eq132}
\end{equation}
Again, to obtain the final nuclear wave function we need to consider two sub
cases: either the final state be neutron-particle-neutron-hole ($%
n_{f}n_{f}^{\prime -1}$) or proton-particle-proton-hole ($p_{f}p_{f}^{\prime
-1}$).{}

\paragraph{Neutron-particle-neutron-hole final state ($n_{f}n_{f}^{\prime -1} $)}

In this case the final nuclear wave function becomes 
\begin{equation}
\left\vert \Psi _{f}\right\rangle =\left[ c_{n_{f}}^{\dagger
}h_{n_{f}^{\prime }}^{\dagger }\right] _{J_{f}M_{f}}\left\vert
HF\right\rangle .  \label{eq133}
\end{equation}%
The transition amplitude becomes%
\begin{eqnarray}
\left\langle n_{f}n_{f}^{\prime -1};J_{f}\left\Vert \beta
_{L}^{+}\right\Vert p_{i}n_{i}^{-1};J_{i}\right\rangle  &=&\delta
_{n_{i}n_{f}^{\prime }}(-1)^{j_{p_{i}}+j_{n_{f}^{\prime }}+J_{i}+L}\hat{J}%
_{i}\hat{J}_{f}\hat{L}\times   \nonumber \\
&&\left\{ 
\begin{array}{ccc}
J_{i} & J_{f} & L \\ 
j_{n_{f}} & j_{p_{i}} & j_{n_{i}}%
\end{array}%
\right\} \mathcal{M}_{L}(p_{i}n_{i}).  \label{eq134}
\end{eqnarray}

\paragraph{Proton-particle-proton-hole final state ($p_{f}p_{f}^{\prime -1}$)}

The final nuclear wave function is 
\begin{equation}
\left\vert \Psi _{f}\right\rangle =\left[ c_{p_{f}}^{\dagger
}h_{^{^{p_{f}^{\prime }}}}^{\dagger }\right] _{J_{f}M_{f}}\left\vert
HF\right\rangle .  \label{eq135}
\end{equation}%
The transition amplitude becomes:
\begin{eqnarray}
\left( p_{f}p_{f}^{^{\prime }-1};J_{f}\left\Vert \beta _{L}^{+}\right\Vert
p_{i}n_{i}^{-1};J_{i}\right)  &=&\delta
_{p_{i}p_{f}}(-1)^{j_{p_{i}}+j_{p_{f}^{\prime }}+J_{f}+1}\hat{J}_{i}\hat{J}%
_{f}\hat{L}\times   \nonumber \\
&&\left\{ 
\begin{array}{ccc}
J_{i} & J_{f} & L \\ 
j_{p_{f}^{\prime }} & j_{n_{i}} & j_{p_{i}}%
\end{array}%
\right\} \mathcal{M}_{L}(n_{i}p_{f}^{\prime }).  \label{eq136}
\end{eqnarray}
The allowed transitions for Fermi $(L=0)$ decay types, in eqs.(\ref{eq126}),
(\ref{eq130}), (\ref{eq134}), and (\ref{eq136}), can be written as \cite%
{Suhonen2019} 
\begin{eqnarray}
\left( n_{f}n_{f}^{\prime -1};J_{f}\left\Vert \beta _{F}^{-}\right\Vert
n_{i}p_{i}^{-1};J_{i}\right) &=&\delta
_{n_{f}n_{i}}(-1)^{j_{n_{i}}+j_{n_{f}^{\prime }}+J_{f}+1}\hat{J}_{i}\hat{J}%
_{f}\left\{ 
\begin{array}{ccc}
J_{i} & J_{f} & 0 \\ 
j_{n_{f}^{\prime }} & j_{p_{i}} & j_{n_{i}}%
\end{array}%
\right\} \mathcal{M}_{L}(p_{i}n_{f}^{\prime })  \notag \\
&=&-\delta _{n_{f}n_{i}}\delta _{J_{i}J_{f}}\delta _{p_{i}n_{f}^{\prime }}%
\hat{J}_{i}\Delta (j_{n_{i}}j_{n_{f}^{\prime }}J_{i}),  \label{eq137}
\end{eqnarray}%
\begin{eqnarray}
\left( p_{_{f}}p_{_{f}}^{\prime -1};J_{f}\left\Vert \beta
_{F}^{-}\right\Vert n_{i}p_{i}^{-1};J_{i}\right) &=&\delta
_{p_{i}p_{f}^{\prime }}(-1)^{j_{n_{i}}+j_{p_{f}^{\prime }}+J_{i}}\hat{J}_{i}%
\hat{J}_{f}\left\{ 
\begin{array}{ccc}
J_{i} & J_{f} & 0 \\ 
j_{p_{f}} & j_{n_{i}} & j_{p_{i}}%
\end{array}%
\right\} \mathcal{M}_{L}(n_{i}p_{f}),  \notag \\
&=&\delta _{p_{i}p_{f}^{\prime }}\delta _{J_{i}J_{f}}\delta _{n_{i}p_{f}}%
\hat{J}_{i}\Delta (j_{p_{i}}j_{p_{f}}J_{i}).  \label{eq138}
\end{eqnarray}%
\begin{eqnarray}
\left( n_{f}n_{f}^{\prime -1};J_{f}\left\Vert \beta _{F}^{+}\right\Vert
p_{i}n_{i}^{-1};J_{i}\right) &=&\delta _{n_{i}n_{f}^{\prime
}}(-1)^{j_{p_{i}}+j_{n_{f}^{\prime }}+J_{i}}\hat{J}_{i}\hat{J}_{f}\left\{ 
\begin{array}{ccc}
J_{i} & J_{f} & 0 \\ 
j_{n_{f}} & j_{p_{i}} & j_{n_{i}}%
\end{array}%
\right\} \mathcal{M}_{L}(p_{i}n_{i}),  \notag \\
&=&\delta _{n_{i}n_{f}^{\prime }}\delta _{J_{i}J_{f}}\delta _{p_{i}n_{i}}%
\hat{J}_{i}\Delta (j_{n_{i}}j_{n_{f}}J_{i}).  \label{eq139}
\end{eqnarray}%
\begin{eqnarray}
\left( p_{f}p_{f}^{\prime -1};J_{f}\left\Vert \beta _{F}^{+}\right\Vert
p_{i}n_{i}^{-1};J_{i}\right) &=&\delta
_{p_{i}p_{f}}(-1)^{j_{p_{i}}+j_{p_{f}^{\prime }}+J_{f}+1}\hat{J}_{i}\hat{J}%
_{f}\left\{ 
\begin{array}{ccc}
J_{i} & J_{f} & 0 \\ 
j_{p_{f}^{\prime }} & j_{n_{i}} & j_{p_{i}}%
\end{array}%
\right\} \mathcal{M}_{L}(n_{i}p_{f}^{\prime }),  \notag \\
&=&-\delta _{p_{i}p_{f}}\delta _{J_{i}J_{f}}\delta _{n_{i}p_{f}^{\prime }}%
\hat{J}_{i}\Delta (j_{p_{i}}j_{p_{f}^{\prime }}J_{i}).  \label{eq140}
\end{eqnarray}%
Here $\hat{j}=\sqrt{2j+1}$, and $\delta _{pn}$ indicates that the
quantum numbers of the proton and neutron orbitals have to be the same. The
symbol $\Delta (j_{1}j_{_{2}}$ $j)$ denotes the \emph{triangular condition}
\begin{equation}
\left\vert j_{1}-j_{2}\right\vert \leq j\leq \left\vert
j_{1}+j_{2}\right\vert .  \label{eq141}
\end{equation}

\subsection{Calculating $\beta ^{-}$-decay strength $ft$ for $^{16}$N}

The Beta-decay equation is
\begin{equation}
_{7}^{16}\text{N}_{9}\rightarrow _{8}^{16}\text{O}_{8}+\beta ^{-}+\bar{\nu}%
_{e},  \label{eq142}
\end{equation}%
We can calculate the $Q$-vlaue
\begin{equation}
Q_{\beta ^{-}}=10.419\text{ MeV}.  \label{eq143}
\end{equation}%
The allowed transition,
\begin{enumerate}
\item $\nu 0p_{\frac{1}{2}}\rightarrow \pi op_{\frac{1}{2}}\Longrightarrow
(\nu 0p_{\frac{1}{2}})^{-1}(\nu 0d_{\frac{5}{2}})$

\item $\nu 0d_{\frac{5}{2}}\rightarrow \pi op_{\frac{1}{2}}\Longrightarrow
0^{+}$

\item $\nu 0d_{\frac{5}{2}}\rightarrow \pi od_{\frac{5}{2}}\Longrightarrow
(\pi 0d_{\frac{5}{2}})(\nu 0p_{\frac{1}{2}})^{-1}$

\item $\nu 0p_{\frac{1}{2}}\rightarrow \pi op_{\frac{1}{2}}$ $\&$ $(\nu 0d_{%
\frac{5}{2}})\rightarrow (\nu 1s_{\frac{1}{2}})\Longrightarrow $ $\nu 1s_{%
\frac{1}{2}}(\nu 0p_{\frac{1}{2}})^{-1}$

\item $\nu 0d_{\frac{5}{2}}\rightarrow \pi 1s_{\frac{1}{2}}\Rightarrow (\pi
1s_{\frac{1}{2}})(\pi 0p_{\frac{1}{2}})^{-1}$

Note that: the selection rule prohibits transition among $d$-states. There
is no transition from $0p_{\frac{1}{2}}\rightarrow 1s_{\frac{1}{2}}$because
the $Q$-value of the decay is smaller than the energy gap.
\end{enumerate}

The initial state is neutron-particle-proton hole ($n_{i}p_{i}^{-1}$) in eq.(%
\ref{eq125-1}). Let the final state be neutron-particle-neutron hole or
proton-particle-proton hole this, 
\begin{equation}
\Psi _{f}=c_{\pi }^{\dagger }h_{\pi }^{\dagger }\left\vert HF\right\rangle
=\left\vert p_{f}p_{f}^{-1}\right\rangle \text{ or }\Psi _{f}=c_{\nu
}^{\dagger }h_{\nu }^{\dagger }\left\vert HF\right\rangle =\left\vert
n_{f}n_{f}^{-1}\right\rangle .  \label{eq144}
\end{equation}%

\begin{figure}[ptbh]
\begin{center}
\includegraphics[scale=2.0]{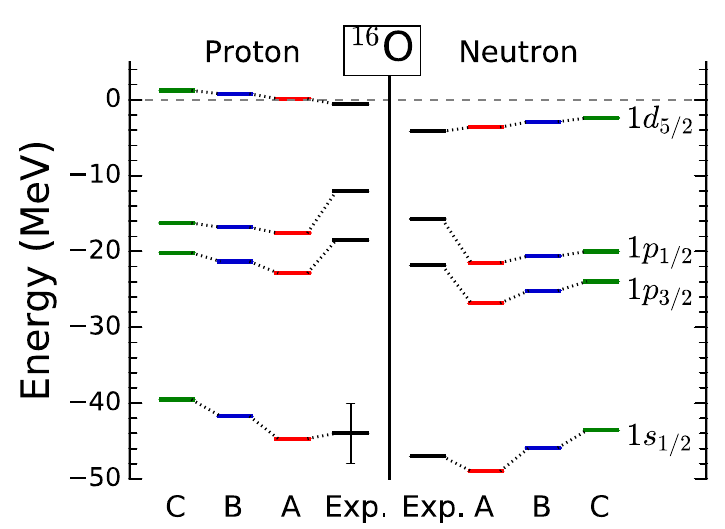}
\end{center}
\par\vspace{-0.7cm}
\caption{Single-particle energies of $^{16}$O calculated
by relativistic Brueckner-Hartree-Fock (RBHF) theory
using the interactions Bonn A, B, and C \cite{Shen2019}, in
comparison with experimental data in \cite{Coraggio2003}.}
\label{fig_o16_sp}
\end{figure}

Let us take the transitions one by one:
\paragraph{1- $\nu 0p_{\frac{1}{2}}\rightarrow (\pi 0p_{\frac{1}{2}})\Longrightarrow
\nu 0d_{\frac{5}{2}}(\nu 0p_{\frac{1}{2}})^{-1}$:} Using the SP states shown
in fig.(\ref{fig_o16_sp}), this transition is energy absorbing transition $%
\Delta E=-4$ MeV. Using eq.(\ref{eq126}), the Gamow-Teller decay amplitude
becomes
\begin{eqnarray}
\left\langle \nu 0d_{\frac{5}{2}}(\nu 0p_{\frac{1}{2}})^{-1};J_{f}\left\Vert
B_{GT}^{-}\right\Vert \nu 0d_{\frac{5}{2}}(\pi 0p_{\frac{1}{2}%
})^{-1}:2^{-}\right\rangle  &=&(-1)^{^{\frac{5}{2}+\frac{1}{2}+J_{f}+1}}%
\hat{J}_{f}\times   \nonumber \\
\sqrt{2\times 1+1}\sqrt{2\times 1+1}\left\{ 
\begin{array}{ccc}
2 & J_{f} & 1 \\ 
\frac{1}{2} & \frac{1}{2} & \frac{5}{2}%
\end{array}%
\right\} \mathcal{M}_{L}(0p_{\frac{1}{2}} &\rightarrow &0p_{\frac{1}{2}}),
\label{eq145} \\
&=&(-1)^{J_{f}}\sqrt{15}\hat{J}_{f}\left\{ 
\begin{array}{ccc}
2 & J_{f} & 1 \\ 
\frac{1}{2} & \frac{1}{2} & \frac{5}{2}%
\end{array}%
\right\} (-\frac{\sqrt{2}}{3}),  \nonumber \\
&=&(-1)^{J_{f}+1}\sqrt{\frac{10}{3}}\hat{J}_{f}\left\{ 
\begin{array}{ccc}
2 & J_{f} & 1 \\ 
\frac{1}{2} & \frac{1}{2} & \frac{5}{2}%
\end{array}%
\right\} ,  \nonumber \\
&=&A_{1(J_{f})}.  \label{eq145_1}
\end{eqnarray}
The possible values of $J_{f}$ can be obtained using the triangular
condition 
\begin{equation*}
(\frac{5}{2}\frac{1}{2}J_{f})\Rightarrow J_{f}=\left\{ 1,2,3\right\} .
\end{equation*}%
The $J_{f}=1$ amplitude returns null because of the $6j$-symbol in eq.(\ref%
{eq145}). Only $J_{f}=\left\{ 2,3\right\} $ are allowed. Thus%
\begin{equation}
\left( \nu 0d_{\frac{5}{2}}(\nu 0p_{\frac{1}{2}})^{-1};J_{f}\left\Vert \beta
_{GT}^{-}\right\Vert \nu 0d_{\frac{5}{2}}(\pi 0p_{\frac{1}{2}%
})^{-1}:2^{-}\right) =A_{1}(J_{f}).  \label{eq146}
\end{equation}%
Performing the calculation we find%
\begin{equation*}
A_{1}(2)=0.608581\text{ and }A_{1}(3)=1.13855.
\end{equation*}%
For Fermi transition, we must have $J_{f}=J_{i}$ according to the selection
rule. Thus the Fermi decay amplitude becomes%
\begin{equation}
\left( \nu 0d_{\frac{5}{2}}(\nu 0p_{\frac{1}{2}})^{-1};2^{-}\left\Vert \beta
_{F}^{-}\right\Vert \nu 0d_{\frac{5}{2}}(\pi 0p_{\frac{1}{2}%
})^{-1}:2^{-}\right) =-\sqrt{5}  \label{eq147}
\end{equation}

\paragraph{2- $\nu 0d_{\frac{5}{2}}\rightarrow (\pi 0d_{\frac{5}{2}})\Rightarrow \pi
0d_{\frac{5}{2}}(\pi 0p_{_{\frac{1}{2}}})^{-1}$}: This is energy absorbing
transition $\Delta E=-3.5$ MeV. Using eq.(\ref{eq130}), the GT decay
amplitude becomes, 
\begin{eqnarray}
\left\langle \pi 0d_{\frac{5}{2}}(\pi 0p_{\frac{1}{2}})^{-1};J_{f}\left\Vert
B_{GT}\right\Vert \nu 0d_{\frac{5}{2}}(\pi 0p_{\frac{1}{2}%
})^{-1};2^{-}\right\rangle  &=&(-1)^{^{\frac{5}{2}+\frac{1}{2}+2+1}}\hat{J}_{f}\times  
\nonumber \\
\sqrt{2\times 2+1}\sqrt{2\times 1+1}\left\{ 
\begin{array}{ccc}
2 & J_{f} & 1 \\ 
\frac{5}{2} & \frac{5}{2} & \frac{1}{2}%
\end{array}%
\right\} \mathcal{M}_{L}(d_{\frac{5}{2}} &\rightarrow &d_{\frac{5}{2}}),
\label{eq148} \\
&=&\sqrt{42}\hat{J}_{f}\times   \nonumber \\
&&\left\{ 
\begin{array}{ccc}
2 & J_{f} & 1 \\ 
\frac{5}{2} & \frac{5}{2} & \frac{1}{2}%
\end{array}%
\right\} ,  \nonumber \\
&=&A_{2}(J_{f}).  \label{eq149}
\end{eqnarray}
For Fermi transition, we must have $J_{f}=J_{i}$ according to the selection
rule. Thus the Fermi decay amplitude becomes%
\begin{equation}
\left\langle \pi 0d_{\frac{5}{2}}(\pi 0p_{\frac{1}{2}})^{-1};2^{-}\left\Vert \beta
_{F}^{-}\right\Vert \nu 0d_{\frac{5}{2}}(\pi 0p_{\frac{1}{2}%
})^{-1};2^{-}\right\rangle =+\sqrt{5}  \label{eq150}
\end{equation}

\paragraph{3- The transition $\nu 0d_{\frac{5}{2}}\rightarrow \pi 0p_{_{\frac{1}{2}%
}}\Rightarrow 0^{+}$}: yields $\left\langle HF;0^{+}\left\Vert \beta
_{GT}\right\Vert \nu 0d_{\frac{5}{2}}(\pi 0p_{\frac{1}{2}})^{-1}:2^{-}%
\right\rangle$ is not explainable using particle-hole theory.

\paragraph{4- $\nu 0p_{\frac{1}{2}}\rightarrow \pi 0p_{\frac{1}{2}}\ $and $\nu 0d_{%
\frac{1}{2}}\rightarrow \nu 1s_{\frac{1}{2}}\Rightarrow \nu 1s_{\frac{1}{2}%
}(\nu 0p_{\frac{1}{2}})^{-1}$:} Using eq.(\ref{eq126}) the GT decay
amplitude becomes, 
\begin{equation}
\left\langle \nu 1s_{\frac{1}{2}}(\nu 0p_{\frac{1}{2}})^{-1};J_{f}\left\Vert \beta
_{GT}\right\Vert \nu 0d_{\frac{5}{2}}(\pi 0p_{\frac{1}{2}})^{-1};2\right\rangle=0,
\label{eq151}
\end{equation}%
because neutron final particle $(\nu 1s_{\frac{1}{2}})$ does not match the
neutron in initial particle $(\nu 0d_{\frac{5}{2}})$.

\paragraph{5- $\nu 0d_{\frac{5}{2}}\rightarrow \pi 1s_{\frac{1}{2}}\Rightarrow (\pi
1s_{\frac{1}{2}})(\pi 0p_{\frac{1}{2}})^{-1}$:} Using eq.(\ref{eq130}), for
GT transition, the GT decay amplitude becomes 
\begin{equation}
\left\langle \pi 1s_{\frac{1}{2}}(\pi 0p_{\frac{1}{2}})^{-1};J_{f}\left\Vert \beta
_{GT}\right\Vert \nu 0d_{\frac{5}{2}}(\pi 0p_{\frac{1}{2}})^{-1};2\right\rangle =0,
\label{eq152}
\end{equation}%
because Initial and final orbital angular momenta do not match.

Using eq.(\ref{eq146}) and eq.(\ref{eq149}),
\begin{eqnarray}
\left\langle 2^{-}\right\vert \beta _{GT}\left\vert 2_{gs}^{-}\right\rangle
&=&\frac{A_{1}(2)+A_{2}(2)}{\sqrt{2}}=\frac{0.608581+2.55604}{\sqrt{2}}%
=2.23772,  \label{eq153} \\
\left\langle 3^{-}\right\vert \beta _{GT}\left\vert 2_{gs}^{-}\right\rangle
&=&\frac{A_{1}(3)+A_{2}(3)}{\sqrt{2}}=\frac{(1.13855+0.57735)}{\sqrt{2}}%
=1.28812.  \notag
\end{eqnarray}%
Make use eq.(\ref{eq153}) and eq.(\ref{eq70}),obtain
\begin{eqnarray}
B_{GT}\left\langle 2_{gs}^{-}\rightarrow 2^{-}\right\rangle  &=&\frac{%
g_{^{2}A}}{2J_{i}+1}\left\vert \left\langle 2^{-}\right\vert \beta
_{GT}\left\vert 2_{gs}^{-}\right\rangle \right\vert ^{2}=1.56481,
\label{eq154} \\
B_{GT}\left\langle 2_{gs}^{-}\rightarrow 3^{-}\right\rangle  &=&\frac{%
g_{^{2}A}}{2J_{i}+1}\left\vert \left\langle 3^{-}\right\vert \beta
_{GT}\left\vert 2_{gs}^{-}\right\rangle \right\vert ^{2}=0.518517.  \notag
\end{eqnarray}%
Make use eq.(\ref{eq154}) into eq.(\ref{eq66}),eq.(\ref{eq72}), we obtain,
\begin{eqnarray}
\log ft(2_{gs}^{-} &\rightarrow &2^{-})=3.59\exp 4.3,  \label{eq155}
\\
\log ft(2_{gs}^{-} &\rightarrow &3)=4.074\exp 4.5.  \notag
\end{eqnarray}

\subsection{ Calculating $\beta ^{+}$/EC decay strength function $ft$ for $^{40}$Sc}

The $\beta ^{+}/$ EC equation is
\begin{equation}
_{21}^{40}\text{Sc}_{19}+e^{-}\rightarrow _{20}^{40}\text{Ca}_{20}+\nu _{e}.
\label{eq156}
\end{equation}
the calculated $Q$-vlaue is
\begin{equation}
Q_{EC}=14.320\text{ MeV}.  \label{eq157}
\end{equation}

\begin{figure}[ptbh]
\begin{center}
\includegraphics[scale=1.75]{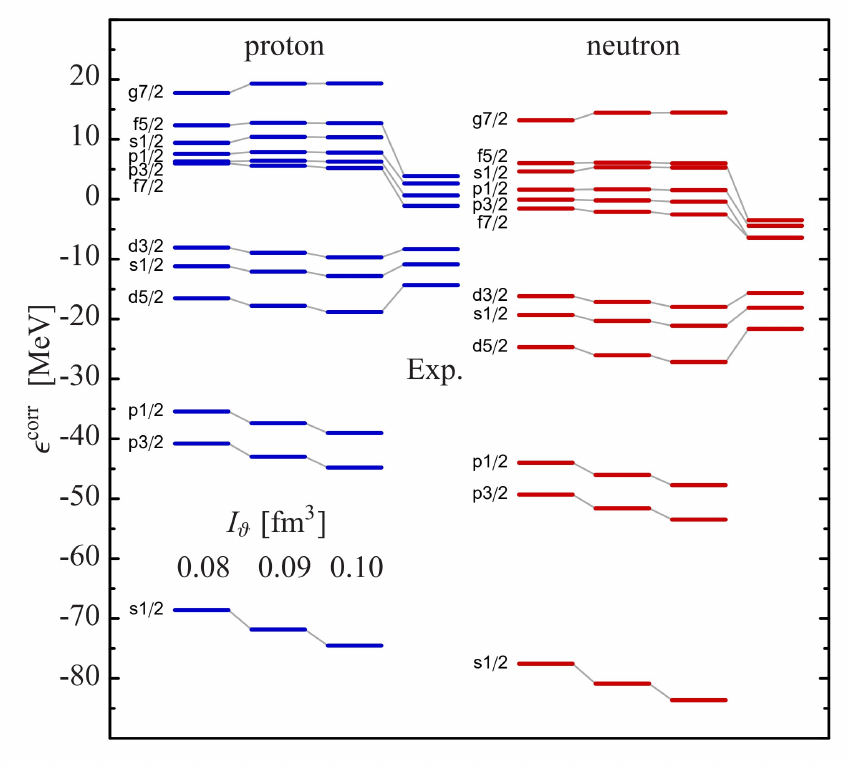}
\end{center}
\par\vspace{-0.7cm}
\caption{Single-particle energies of $^{40}$Ca calculated by
Hartree-Fock (RBHF) theory using the interactions AV18 \cite{Papa2005},
in comparison with experimental data from \cite{Isakov2002}.}
\label{fig_ca40_sp}
\end{figure}

According to the SP states of $^{40}$Ca shown in fig.(\ref{fig_ca40_sp}) and
using eq.(\ref{eq133}) initial nuclear state, 
\begin{equation}
\left\vert\Psi _{i}\right\rangle=\left\vert p_{i}n_{i}^{-1}\right\rangle =\left\vert \pi 0f_{\frac{7%
}{2}}(\nu 0d_{\frac{3}{2}})^{-1}\right\rangle ,  \label{eq158}
\end{equation}%
whereas the final nuclear state%
\begin{equation}
\left\vert\Psi _{f}\right\rangle=c_{\pi }^{\dagger }h_{\pi }^{\dagger }\left\vert HF\right\rangle
=\left\vert p_{f}p_{f}^{-1}\right\rangle \text{ or }\Psi _{f}=c_{\nu
}^{\dagger }h_{\nu }^{\dagger }\left\vert HF\right\rangle =\left\vert
n_{f}n_{f}^{-1}\right\rangle .  \label{eq159}
\end{equation}%
We have the following possible SP transitions:
\paragraph{1- $\pi 0d_{\frac{3}{2}}(-6.2$ MeV$)\rightarrow \nu 0d_{\frac{3}{2}%
}(-14.2$ MeV$)\Rightarrow \pi 0f_{\frac{7}{2}}(\pi 0d_{\frac{3}{2}})^{-1}$:}
Thus $\Delta E=8$ MeV The GT decay amplitude using eq.(\ref{eq136}) becomes
\begin{eqnarray}
\left\langle \pi 0f_{\frac{7}{2}}(\pi 0d_{\frac{3}{2}})^{-1};J_{f}\left\Vert
\beta _{L}^{+}\right\Vert \pi 0f_{\frac{7}{2}}(\nu 0d_{\frac{3}{2}%
})^{-1};4\right\rangle  &=&(-1)^{\frac{7}{2}+\frac{3}{2}+J_{f}+1}\times  
\nonumber \\
\sqrt{2\times 4+1}\hat{J}_{f}\sqrt{2\times 1+1}\left\{ 
\begin{array}{ccc}
4 & J_{f} & 1 \\ 
\frac{3}{2} & \frac{3}{2} & \frac{7}{2}%
\end{array}%
\right\} \mathcal{M}_{L}(0d_{\frac{3}{2}} &\rightarrow &0d_{\frac{3}{2}})
\label{eq160} \\
&=&(-1)^{J_{f}}\sqrt{9}\sqrt{3}J_{f}\ \times   \nonumber \\
&&\left\{ 
\begin{array}{ccc}
4 & J_{f} & 1 \\ 
\frac{3}{2} & \frac{3}{2} & \frac{7}{2}%
\end{array}%
\right\} (-\frac{2}{\sqrt{5}})  \nonumber \\
&=&(-1)^{J_{f}}(-6\sqrt{\frac{3}{5}})\hat{J}_{f}\left\{ 
\begin{array}{ccc}
4 & J_{f} & 1 \\ 
\frac{3}{2} & \frac{3}{2} & \frac{7}{2}%
\end{array}%
\right\}   \nonumber \\
&=&A_{1(J_{f})}.  \nonumber
\end{eqnarray}
Only $J_{f}=\left\{ 3,4,5\right\} $ are allowed. $A_{1}(3)=-1.54919$, $%
A_{1}(4)=-1.07331$, and $A_{1}(5)=1.3594$. Make use of eq.(\ref{eq160}) into
eq.(\ref{eq70}), we obtain
\begin{eqnarray}
B_{GT}(4^{-}\rightarrow 3) &=&\frac{g_{A}^{2}}{2J_{i}+1}\left\vert \langle
^{-}4\right\vert B_{GT}\left\vert 3\rangle \right\vert ^{2}  \nonumber \\
&=&\frac{(1.25)^{2}}{2\times 4+1}(-1.54919)^{2}=0.416665\label{eq161}
\end{eqnarray}
\begin{equation}
B_{GT}(4^{-}\rightarrow 4)=\frac{g_{A}^{2}}{2J_{i}+1}\left\vert \langle
^{-}4\right\vert B_{GT}\left\vert 4\rangle \right\vert ^{2}=0.199999
\label{eq162}
\end{equation}

For Fermi transition, the Fermi decay amplitude becomes,
\begin{eqnarray}
\left\langle \pi 0f_{\frac{7}{2}}(\pi 0d_{\frac{3}{2}})^{-1};J_{f}\left\Vert
\beta _{L}^{+}\right\Vert \pi 0f_{\frac{7}{2}}(\nu 0d_{\frac{3}{2}%
})^{-1};4\right\rangle  &=&(-1)^{^{\frac{7}{2}+\frac{3}{2}+J_{f}+1}}\times  
\nonumber \\
\sqrt{2\times 4+1}\hat{J}_{f}\sqrt{2\times 0+1}\left\{ 
\begin{array}{ccc}
4 & J_{f} & 0 \\ 
\frac{3}{2} & \frac{3}{2} & \frac{7}{2}%
\end{array}%
\right\} \mathcal{M}_{0}(0d_{\frac{3}{2}} &\rightarrow &0d_{\frac{3}{2}})
\label{eq163} \\
&=&9(-1)^{4}\left\{ 
\begin{array}{ccc}
4 & J_{f} & 0 \\ 
\frac{3}{2} & \frac{3}{2} & \frac{7}{2}%
\end{array}%
\right\} (2)  \nonumber \\
&=&-3.  \nonumber
\end{eqnarray}
Only $J_{f}=4$ contributes to the amplitude (\ref{eq163})
\begin{equation*}
\langle 4^{-}\left\vert \beta _{F}\right\vert 4\rangle=-3.
\end{equation*}%
Using eq.(\ref{eq163}) into eq.(\ref{eq69})
\begin{equation}
B_{F}=\frac{(1.0)^{2}}{2\times 4+1}(-3)^{2}=1.  \label{eq164}
\end{equation}%
Make use of eq.(\ref{eq160}) into eq.(\ref{eq70}), we obtain
\begin{equation}
B_{GT}(4^{-}\rightarrow 5)=\frac{g_{A}^{2}}{2J_{i}+1}\left\vert \langle
^{-}4\right\vert \beta _{GT}\left\vert 5\rangle \right\vert ^{2}=0.320828.
\label{eq165}
\end{equation}%
Make use eq.(\ref{eq161}) \ and eq.(\ref{eq162}) and eq.(\ref{eq164}) and
eq. (\ref{eq165})into eq.(\ref{eq66}) ,and eq.(\ref{eq72}), we obtain,
$\log $ $ft$ $(4^{-}\rightarrow 3)=4.16$ experimental value is 4.8

$\log $ $ft$ $(4^{-}\rightarrow 4)=3.70$ experimental value is 4.6

$\log $ $ft$ $\ (4^{-}\rightarrow 5)=4.28$ experimental value is 4.7.

\paragraph{2- $\pi 0d_{\frac{5}{2}}(-12.0$ MeV$)\rightarrow \nu 0d_{\frac{3}{2}%
}(-14.2$ MeV $)\Rightarrow \pi 0f_{\frac{7}{2}}(\pi 0d_{\frac{5}{2}})^{-1}$:}
Thus $\Delta E=2.2$ MeV Using eq.(\ref{eq136}), the Gamow-Teller decay
amplitude becomes
\begin{eqnarray}
\left\langle \pi 0f_{\frac{7}{2}}(\pi 0d_{\frac{5}{2}})^{-1};J_{f}\left\Vert
\beta _{L}^{+}\right\Vert \pi 0f_{\frac{7}{2}}(\nu 0d_{\frac{3}{2}%
})^{-1};4\right\rangle  &=&  \nonumber \\
(1)^{J_{f}+1}\sqrt{2\times 4+1}\hat{J}_{f}\sqrt{2\times 1+1} &&\left\{ 
\begin{array}{ccc}
4 & J_{f} & 1 \\ 
\frac{5}{2} & \frac{3}{2} & \frac{7}{2}%
\end{array}%
\right\} \frac{4}{\sqrt{5}},  \label{eq166} \\
&=&(1)^{J_{f}+1}(-12\sqrt{\frac{3}{5}})J_{f}\left\{ 
\begin{array}{ccc}
4 & J_{f} & 1 \\ 
\frac{5}{2} & \frac{3}{2} & \frac{7}{2}%
\end{array}%
\right\} ,  \nonumber \\
&=&A_{2(J_{f})}  \nonumber
\end{eqnarray}
only $J_{f}=\left\{ 3,4,5\right\} $, are allowed $A_{2}(3)=$ $1.34164$ and $%
A_{2}(4)=2.66983$ and $A_{2}(5)=3.55978$. Use eq.(\ref{eq166}) and eq.(\ref%
{eq70}) we get
\begin{equation}
B_{GT}(4^{-}\rightarrow 3)=\frac{g_{A}^{2}}{2J_{i}+1}\left\vert \langle
^{-}4\right\vert B_{GT}\left\vert 3\rangle \right\vert ^{2}=\frac{%
(1.25)^{2}}{2\times 4+1}(1.34164)^{2}=0.3125,  \label{eq167}
\end{equation}
\begin{equation}
B_{GT}(4^{-}\rightarrow 4)=\frac{g_{A}^{2}}{2J_{i}+1}\left\vert \langle
^{-}4\right\vert B_{GT}\left\vert 4\rangle \right\vert ^{2}=\frac{(1.25)^{2}%
}{2\times 4+1}(2.66983)^{2}=1.2375,  \label{eq168}
\end{equation}
\begin{equation}
B_{GT}(4^{-}\rightarrow 5)=\frac{g_{A}^{2}}{2J_{i}+1}\left\vert \langle
^{-}4\right\vert B_{GT}\left\vert 5\rangle \right\vert ^{2}=\frac{(1.25)^{2}%
}{2\times 4+1}(3.55978)^{2}=2.20001.  \label{eq169}
\end{equation}%
The Fermi transition is zero for this transition for any value of \thinspace
\thinspace $J_{f}$. Use eq.(\ref{eq167}), eq.(\ref{eq168}), and eq. (\ref%
{eq169}) into eq.(\ref{eq66}) and eq.(\ref{eq72}) , we obtain,

$\log $ $ft(4^{-}\rightarrow 3)=4.29$ experimental value 5.1

$\log $ $ft(4^{-}\rightarrow 4)=3.69$ experimental value 3.3

$\log $ $ft(4^{-}\rightarrow 5)=3.44$ experimental value 4.7 (doesn't exist
experimentally for such energy range).

Complete results are shown in table (\ref{tab_logft_sc40_ca40_EC}).

\begin{table}[htb]
\caption{The calculated strength function $\log ft$ compared with the
experimental values for the $^{40}$Sc to $^{40}$Ca $\beta^+$/EC.}
\label{tab_logft_sc40_ca40_EC}
\par
\begin{center}
\begin{tabular}{lllll}\hline
Transition & $\log ft$ & energy (MeV) & Isospin $T$ & $\log (ft)_{\exp }$ \\ \hline \hline
$\left\vert \langle 4^{-}\right\vert \beta _{GT}\left\vert 3\rangle
\right\vert $ & $4.16$ & $3.73$ & $0$ & $4.8$ \\ 
$\left\vert \langle 4^{-}\right\vert \beta _{GT}\left\vert 4\rangle
\right\vert $ & $3.70$ & $5.61$ & $0$ & $4.6$ \\ 
$\left\vert \langle 4^{-}\right\vert \beta _{GT}\left\vert 5\rangle
\right\vert $ & $4.28$ & $4.49$ & $0$ & $4.7$ \\ 
$\left\vert \langle 4^{-}\right\vert \beta _{GT}\left\vert 3\rangle
\right\vert $ & $4.29$ & $6.58$ &  & $5.1$ \\ 
$\left\vert \langle 4^{-}\right\vert \beta _{GT}\left\vert 4\rangle
\right\vert $ & $3.69$ & $7.65$ & $1$ & $3.3$ \\ 
$\left\vert \langle 4^{-}\right\vert \beta _{GT}\left\vert 5\rangle
\right\vert $ & $4.44$ & $4.49$ & $0$ & $4.7$\\ \hline
\end{tabular}
\end{center}
\end{table}

\section{Conlusion}

We summarize the results of $\log (ft)$ calculated using particle-hole
theory in table (\ref{tab_logft_Ni16_beta_data}) for $^{16}$Ni decay and
table (\ref{tab_logft_Sc40_EC_data}) for $^{40}$Sc EC.  
\begin{table}[tbh]
\caption{Summary of the calculated $\beta ^{-}$-decay logarithm of
the strength function $\log ft$ for all allowed transitions, using
particle-hole theory for the decay $^{16}$N$\rightarrow ^{16}$O$+e^{-}+\bar{%
\nu}_{e}$ which has $Q_{\beta }=10.419$ MeV. The temperature
calculation is based on hot Thomas-Fermi model \cite{Rashdan_1991}.}
\label{tab_logft_Ni16_beta_data}
\par
\begin{center}
\begin{tabular}{ccccccccc}
\hline
Single Particle                                              & SP transition & Nuclear    & Nuclear state & Isospin & $\log ft$ & $\log ft$ & Temperature 				& $g_{A}$(exp) \\
Transition                                                   & Energy (MeV)  & transition & energy (MeV)  & $T$     & Theory    &  exp      & (MeV) \cite{Rashdan_1991} &               \\ \hline
$\nu 0p_{\frac{1}{2}}\rightarrow \pi 0p_{\frac{1}{2}}$    &               &            &               &         &           &           &                           &\\
$\nu 0d_{\frac{5}{2}}(\nu 0p_{\frac{1}{2}})^{-1}$         & -4   & $ \langle 2^{-}|\beta _{GT}|2^{-}\rangle  $ & 8.87 & 0 & 3.59 & 4.3 & 1.8347 & 0.55464 \\ \\
$\nu 0d_{\frac{5}{2}}\rightarrow \pi 0d_{\frac{5}{2}}$    &      &                                                                 &      &   &       &     &        &\\
$\pi 0d_{\frac{5}{2}}(\pi 0p_{_{\frac{1}{2}}})^{-1}$      & -3.5 & $ \langle3^{-}|\beta _{GT}|2^{-}\rangle   $ & 6.13 & 0 & 4.074 & 4.5 & 1.5135 & 0.76535 \\ \hline
\end{tabular}
\end{center}
\end{table}
\begin{table}[tbh]
\caption{Summary of the calculated $\beta ^{+}/EC$-decay logarithm
of the strength function $\log ft$ for all allowed transitions, using
particle-hole theory for the decay $^{40}$Sc$+e^{-}\rightarrow ^{40}$Ca$+%
\nu _{e}$ which has $Q_{EC}=14.32$ MeV. The temperature calculation
is based on hot Thomas-Fermi model \cite{Rashdan_1991}.}
\label{tab_logft_Sc40_EC_data}
\par
\begin{center}
\begin{tabular}{ccccccccc}
\hline
Single Particle                                              & SP transition & Nuclear    & Nuclear state & Isospin & $\log ft$ & $\log ft$ & Temperature 				& $g_{A}$(exp) \\
Transition                                                   & Energy (MeV)  & transition & energy (MeV)  & $T$     & Theory    &  exp      & (MeV) \cite{Rashdan_1991} &               \\ \hline
$\pi 0d_{\frac{3}{2}}\rightarrow \nu 0d_{\frac{3}{2}}$       &               &            &               &         &           &           &                           &\\
$\pi 0f_{\frac{7}{2}}(\pi 0d_{\frac{3}{2}})^{-1}$ & 8 & $ \langle
3^{-}| \beta _{GT}| 4^{-}\rangle $ & 3.73 & 0
& 4.16 & 4.8 & 0.847 & 0.604433 \\ \\

$\pi 0d_{\frac{5}{2}}\rightarrow \nu 0d_{\frac{3}{2}}$       &               &            &               &         &           &           &                           &\\
$\pi 0f_{\frac{7}{2}}(\pi 0d_{\frac{5}{2}})^{-1}$ & 2.2 & $ \langle
3^{-}| \beta _{GT}| 4^{-}\rangle  $ & 6.58 & ?
& 4.29 & 5.1 & 1.090 & 0.494102 \\ \\

$\pi 0d_{\frac{3}{2}}\rightarrow \nu 0d_{\frac{3}{2}}$       &               &            &               &         &           &           &                           &\\
$\pi 0f_{\frac{7}{2}}(\pi 0d_{\frac{3}{2}})^{-1}$ & 8 & $ \langle
4^{-}| \beta _{GT}| 4^{-}\rangle  $ & 5.61 & 0
& 3.70 & 4.6 & 1.013 & 1.09832 \\ \\

$\pi 0d_{\frac{5}{2}}\rightarrow \nu 0d_{\frac{3}{2}}$       &               &            &               &         &           &           &                           &\\
$\pi 0f_{\frac{7}{2}}(\pi 0d_{\frac{5}{2}})^{-1}$ & 2.2 & $ \langle
4^{-}| \beta _{GT}| 4^{-}\rangle  $ & 7.65 & 1
& 3.69 & 3.3 & 1.169 & 1.97228 \\ \\

$\pi 0d_{\frac{3}{2}}\rightarrow \nu 0d_{\frac{3}{2}}$       &               &            &               &         &           &           &                           &\\
$\pi 0f_{\frac{7}{2}}(\pi 0d_{\frac{3}{2}})^{-1}$ & 8 & $ \langle
5^{-}| \beta _{GT}| 4^{-}\rangle  $ & 4.49 & 0
& 4.28 & 4.7 & 0.918 & 0.772868 \\ \\

$\pi 0d_{\frac{5}{2}}\rightarrow \nu 0d_{\frac{3}{2}}$       &               &            &               &         &           &           &                           &\\
$\pi 0f_{\frac{7}{2}}(\pi 0d_{\frac{5}{2}})^{-1}$ & 2.2 & $ \langle
5^{-}| \beta _{GT}| 4^{-}\rangle  $ & 4.49 & 0
& 4.44 & 4.7 & 0.918 & 0.295141 \\ \hline
\end{tabular}%
\end{center}
\end{table}

We utilize the one-particle-hole theory to calculate the half-lives of odd-even nuclei of \ce{^{15}O}, \ce{^{17}F}, \ce{^{39}Ca}, and \ce{^{41}Sc}. By comparing the calculated values with the experimental data we found discrepancies ranging from 27.4\% for \ce{^{39}Ca} to 41.4\% for \ce{^{17}F}. The experimental half-lives always exceed those of the theoretical values. This is attributed to the fact that the one-particle-hole theory is dependent on the SP states and doesn’t account for the residual force among the even valence nucleons. This suppresses the SP transitions in the initial and final nuclear wave functions $\Psi_i$ and $\Psi_f$. The simplicity of this model makes it a favorable technique to obtain a rough estimate for the half-lives of the $\beta$-decays

In the two-particle-hole scheme, we calculate the strength function $\log ft$ of the $\beta^+$/EC decay of the even-even nucleus \ce{^{56}Ni}. The value is $\log ft=2.46$. The experimental value is 4.4 which exceeds the calculated value $\log ft=4.4$. A 44.1\% discrepancy is an indication of the failure of the particle-hole theory. Again, as shown in table (\ref{tab_beta_class}), the transition is unfavoured allowed, meaning that the single-particle transition is suppressed during the initial and the final states due to the residual two-body interaction.

For odd-odd nuclei, \ce{^{16}N} and \ce{^{40}Sc}, the results take different turn. Except for $\pi 0d_{\frac{5}{2}}\rightarrow \nu 0d_{\frac{3}{2}}$ transition, corresponds to
$\langle 4^{-}|\beta _{GT}|4^{-}\rangle$ amplitude, for the \ce{^{40}Sc}, all experimental values of the strength functions are less than those of the experimental ones. The effect of the NN residual force is diluted for the odd-odd nuclei. The discrepancies here are merely due to the error in determining the exact energy levels of the SP states. The theory is a success in this situation.

\begin{figure}[ht]
\begin{center}
\includegraphics[scale=0.75]{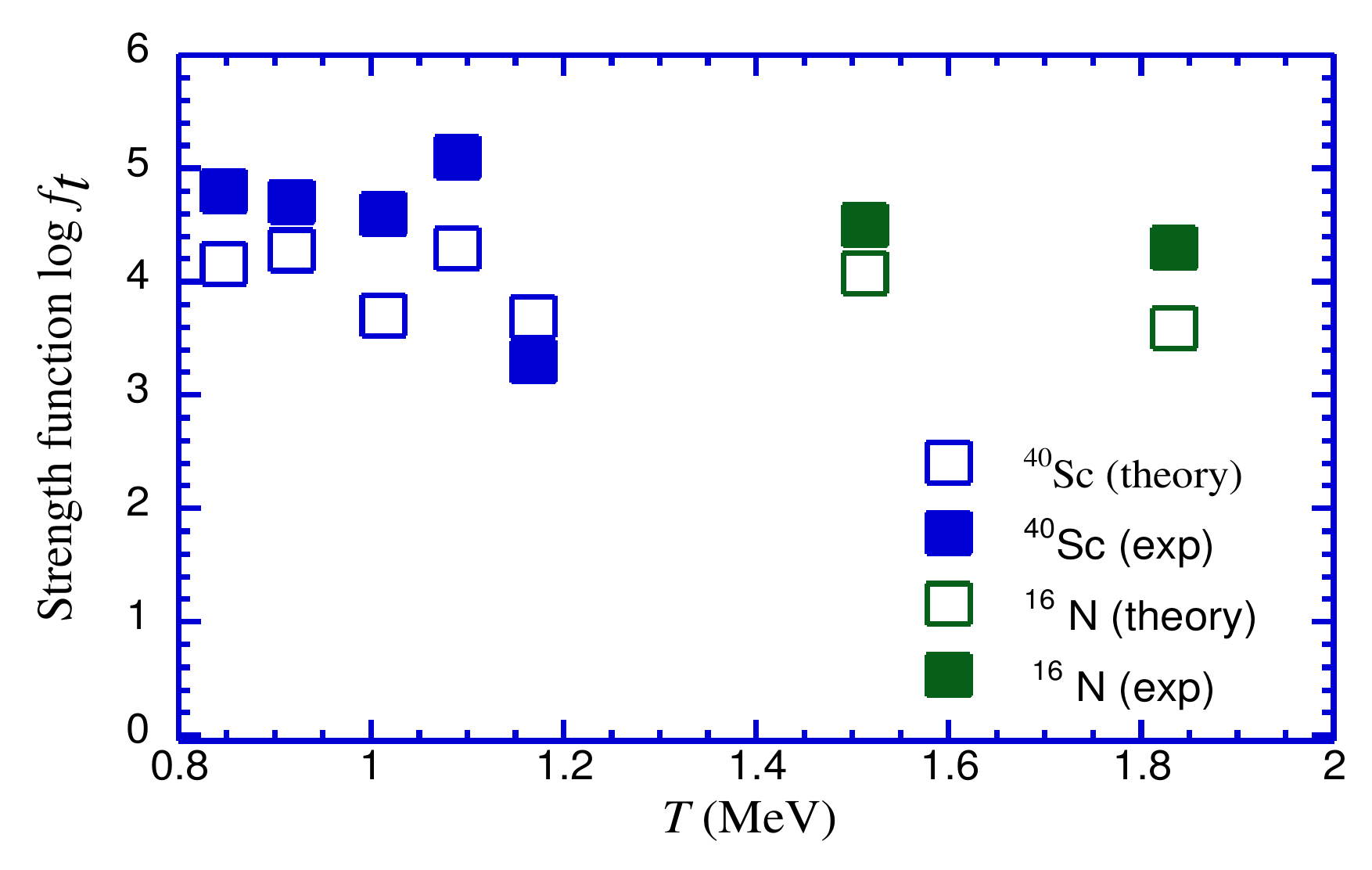}
\end{center}
\par\vspace{-0.7cm}
\caption{The temperature-dependent of the theoretical and experimental strength function $\log ft$ for \ce{^{40}Sc} and \ce{^{16}N} (see colored legends).}
\label{figlogftvst}
\end{figure}

The temperature-dependent of the $\log ft$ is shown in fig.(\ref{figlogftvst}). The general trend is that the value of $log ft$ is slowly decreasing with temperature. The fluctuations in the values reflect the dependent on the shell configurations used to compute the amplitude. The amplitude $\langle3^{-}|\beta _{GT}|2^{-}\rangle$ for \ce{^{16}N} and
$\langle 5^{-}|\beta _{GT}|4^{-}\rangle$ for \ce{^{40}Sc} have the closer values to the experimental ones. The drop of the $\log ft$ value for $\langle 4^{-}|\beta _{GT}|4^{-}\rangle$ amplitude is attributed to the fact that Gamow-Teller transition is more likely to occur for $\Delta J=0$.  

We need to go more steps further and use two-particle-two-hole configurations to “quench” the Gamow-Teller strengths \cite{PhysRevC.105.014321} and implement quasi-particle random phase approximation \cite{PhysRevC.101.044305} for more accurate computations.

\clearpage
\bibliography{atbibilio}{}

\end{document}